\documentclass[aps,prfluids,12pt,superscriptaddress,showkeys,longbibliography,floatfix]{revtex4-2}

\usepackage{amsmath,amssymb,bm}
\usepackage{graphicx}
\usepackage{booktabs}
\usepackage{siunitx}
\usepackage{xcolor}
\usepackage{hyperref}
\hypersetup{colorlinks=true,linkcolor=blue,citecolor=blue,urlcolor=blue}

\newcommand{\Er}{E_r}
\newcommand{\kb}{k_B}
\newcommand{\QD}{Q_D}
\newcommand{\Qmu}{Q_\mu}
\newcommand{\RCS}{\Sigma_{\rm RCS}}
\newcommand{\SigVSS}{\Sigma_{\rm VSS}}
\newcommand{\RDA}{\chi_{\rm RDA}}
\newcommand{\EPAPS}{\mathrm{EPAPS}}
\newcommand{\NN}{\mathrm{NN}}
\newcommand{\dd}{\mathrm{d}}
\newcommand{\eps}{\varepsilon}

\begin{document}

\title{Transport-preserving neural ab initio scattering kernels for rarefied binary gas mixtures}

\author{Ehsan Roohi}
\email{roohie@umass.edu}
\affiliation{Department of Mechanical and Industrial Engineering, University of Massachusetts Amherst, Amherst, Massachusetts 01003, USA}

\date{\today}

\begin{abstract}
Neural surrogates for molecular scattering provide a route to continuously evaluable and differentiable direct simulation Monte Carlo (DSMC) collision kernels, but a small pointwise deflection-angle error is not sufficient evidence that a learned map is kinetically reliable.  Diffusion, viscosity, representative collision rates, angular redistribution, and mixture relaxation are nonlinear functionals of the same scattering measure.  We therefore develop a multiscale validation framework for neural ab initio scattering kernels that combines angular regression, transport cross sections, Ohr-style representative quantities, cumulative angular measures, Fourier spectral content, impact-grid and angular-noise robustness, loss-ablation diagnostics, and three solver-level DSMC mixture tests.  The framework is demonstrated on a refined argon--argon J\"ager table and on helium--argon ab initio EPAPS data of Sharipov and Benites represented by a neural equal-area scattering surrogate.  For He--Ar over $\Er/\kb\ge10~\mathrm{K}$, the surrogate preserves $\QD$, $\Qmu$, $\Qmu/\QD$, $\RCS$, and $\SigVSS$ within $0.75\%$, $1.37\%$, $0.84\%$, $1.21\%$, and $1.46\%$, respectively.  The cumulative angular measure agrees within $1.43\%$, the median relative $L_2$ error of $\chi(q)$ is $3.4\times10^{-3}$, and the high-mode spectral-energy ratio is essentially unbiased.  The same neural He--Ar kernel is then embedded in periodic DSMC mixture problems that separately probe mass diffusion, momentum diffusion, and two-dimensional field-level mixing.  A sinusoidal composition mode is reproduced over three independent realizations with a mean normalized-history error of $1.28\pm0.22\%$ and $D_{\NN}/D_{\EPAPS}=1.015\pm0.013$.  A transverse shear wave is reproduced with a $1.58\%$ history error and $\nu_{\NN}/\nu_{\EPAPS}=0.989$.  Finally, a two-dimensional periodic Ar--He shear-layer calculation activates finite composition gradients, coherent shear, scalar dissipation, and interspecies slip without wall-model ambiguity; the neural kernel reproduces the EPAPS mixing-index history with a $0.124\%$ error, the selected-time $X_{\rm He}$ field with a $2.56\%$ relative $L_2$ error, and the species-slip field with a $6.37\%$ error.  A Chapman--Enskog analysis from the EPAPS diffusion cross section gives $D_{12}^{CE}=0.916~\mathrm{m^2/s}$ at $300~\mathrm{K}$ and $2\times10^{21}~\mathrm{m^{-3}}$, while the apparent scalar-diffusion ratio extracted from the sheared DSMC field has median $D_{\rm app}^{\NN}/D_{\rm app}^{\EPAPS}=1.003$.  The same diagnostics also identify the low-energy $1$--$100~\mathrm{K}$ interval as the most sensitive part of the map, so surrogate accuracy should be interpreted through the collision-energy and transport weights sampled by the target application.  These results support the central hypothesis that DSMC-ready neural scattering kernels should be validated by preservation of angular measures, transport projections, and mixture dynamics rather than by angle error alone.
\end{abstract}


\maketitle

\section{Introduction}

Rarefied gas dynamics requires a kinetic description when the molecular mean free path becomes comparable with the macroscopic length scale.  This situation arises in high-altitude aerodynamics, hypersonic flight, micro- and nano-scale gas devices, cryogenic low-density flows, and gas-mixture transport.  The direct simulation Monte Carlo (DSMC) method remains the most widely used particle method for such problems because it replaces the Boltzmann collision integral by stochastic binary-collision sampling while retaining molecular free flight between collisions \cite{Bird1963,Bird1970,Nanbu1980,RoohiAkhlaghiStefanov2025DSMCBook}.  Its reliability depends on several numerical limits, e.g., cell size, time step, particle number, and sampling duration, but it also depends fundamentally on the intermolecular collision model.  Hard-sphere, variable-hard-sphere (VHS), and variable-soft-sphere (VSS) models provide efficient phenomenological closures and have been central to practical DSMC for decades \cite{KouraMatsumoto1991}.  Nevertheless, such models reduce a complex interaction potential to a small number of effective parameters.  In regimes where attractive tails, orbiting, quantum-informed potential-energy curves, or cross-species interactions matter, the resulting transport coefficients and angular redistribution can depart from those of the real gas.

A more faithful elastic collision description is obtained from a scattering map.  For a pair of species $i$ and $j$, the relative energy $\Er=m_{ij}g^2/2$ and impact parameter $b$ determine a deflection angle $\chi(\Er,b)$ through the underlying potential.  Once $\chi$ is known, the post-collision relative velocity direction can be sampled and the molecular momentum and energy are updated exactly for an elastic binary encounter.  This viewpoint is attractive because it separates two questions that are often conflated.  The first is the collision-rate closure: how often should candidate pairs be selected in a DSMC cell?  The second is the accepted microscopic scattering law: how should an accepted pair change direction?  For realistic potentials both questions are difficult.  Long-range interactions can make total cross sections sensitive to cutoff conventions, while repeated trajectory integration during a DSMC run can be prohibitive.

Accurate ab initio potentials and transport data provide an increasingly important reference for such collision models.  For dilute gases, high-accuracy pair potentials and transport integrals have been developed for helium, argon, and their mixtures \cite{Aziz1995,Bertoncini1970,Vogel2010,Jager2011}.  Sharipov and co-workers have produced ab initio transport data and scattering-related supplementary tables for noble-gas mixtures, including helium--argon and multicomponent noble gas systems \cite{SharipovBenites2015,SharipovBenites2019,Sharipov2020,Sharipov2024}.  A closely related line of work has already demonstrated that ab initio potentials can be embedded directly in DSMC or Boltzmann-type calculations for transport recovery, Couette and Fourier/heat-transfer configurations, shock waves, orifice flow, and rarefied benchmark problems \cite{SharipovStrapasson2012,SharipovStrapasson2013,StrapassonSharipov2014Heat,SharipovDias2017Shock,Sharipov2022DSMC}.  The novelty sought here is therefore not the existence of ab initio mixture data or the possibility of using such data in DSMC.  Rather, the question is how a learned, smooth, DSMC-ready representation of an ab initio scattering map should be judged before it is substituted for the native table.  These datasets create the possibility of replacing approximate VHS/VSS angular models by tabulated or learned ab initio scattering kernels.  They also create a new validation problem: if a neural network is trained on a table of deflection angles, what must be checked before the network can be trusted as a DSMC collision kernel?

The simplest answer, small mean angle error, is not enough.  The physically relevant quantities are integrals of the scattering map.  The diffusion cross section $\QD$ weights the angular map by $1-\cos\chi$; the viscosity cross section $\Qmu$ weights it by $1-\cos^2\chi$; representative single-deflection methods form nonlinear combinations of these integrals; and the full angular redistribution is encoded by the push-forward of impact area into $\mu=\cos\chi$.  Small pointwise errors may have little effect on one transport projection but a larger effect on another, especially near low-energy turning regions, rainbow-like structures, or high-curvature branches.  Conversely, a visually noisy angular map can still preserve integral transport quantities if the induced measure is correct.  A useful validation strategy must therefore be transport-aware and measure-aware.

A parallel lesson has emerged in data-driven flow reconstruction.  Neural networks trained only with mean-squared error tend to smooth small-scale structures, even when large-scale visual agreement is good.  Recent work on turbulent-flow sensing demonstrated that composite and spectrally informed loss functions, including RMS-amplitude, correlation, gradient, and Fourier-amplitude terms, substantially improve reconstruction fidelity and preserve small-scale energy content \cite{Guastoni2021,Balasubramanian2026}.  The analogy for scattering is direct: the field to be preserved is not a velocity field $u(x,z)$ but the equal-impact-area scattering map $\chi(q,E)$.  Its Fourier content in $q$, its cumulative distribution in angular space, and its transport projections play the same role as energy spectra, RMS statistics, and coherent structures in turbulence reconstruction.  This motivates a validation philosophy that is broader than pointwise regression.

Machine learning has already been used in kinetic theory and rarefied flows at several levels.  Physics-informed neural networks and neural operators have provided general frameworks for learning solution maps and enforcing physical constraints \cite{Raissi2019,Karniadakis2021,Lu2021DeepONet,Kovachki2023}.  Neural networks have been used for fluid mechanics, data-driven turbulence modeling, and accelerated computational fluid dynamics \cite{Brunton2020,Duraisamy2019,Kochkov2021}.  In kinetic theory, learned collision operators and structure-preserving Boltzmann surrogates have been proposed \cite{Xiao2023,Miller2022,Corbetta2023,Ball2025}.  Our recent work has developed DSMC and rarefied-flow surrogates for Maxwellian distributions, shocks, micro-step flows, hypersonic cylinders, and beyond \cite{RoohiAST2026,RoohiPoF2026a,RoohiPoF2026b,RoohiMicro2026}.  

The central hypothesis of the current work is the following: \emph{a neural ab initio scattering surrogate suitable for DSMC must preserve the angular push-forward measure and its transport projections, not merely the pointwise deflection angle, and this preservation must remain visible when the surrogate is embedded in independent mixture-transport dynamics.}  We test this hypothesis through a hierarchical protocol.  At the angular level, we compare $\chi(q,E)$ directly.  At the transport level, we compare $\QD$, $\Qmu$, and $\Qmu/\QD$.  At the representative-collision level, we evaluate the Ohr-style representative collision cross section $\RCS$, representative deflection angle $\RDA$, and equivalent VSS parameter \cite{Ohr2023}.  At the measure level, we compare cumulative angular distributions $\Sigma(\mu;E)$.  At the spectral level, we compare Fourier amplitudes of the scattering map in the equal-area coordinate.  We then probe robustness to finite impact-area resolution and synthetic angular perturbations.  Finally, to connect these kernel diagnostics to actual DSMC transport response, we solve three periodic binary Ar--He problems.  The first two are single-mode relaxation tests: a sinusoidal composition mode probes mutual diffusion, and a transverse shear wave probes momentum diffusion.  The third is a two-dimensional periodic shear-layer mixing problem that simultaneously contains finite composition gradients, coherent shear, scalar dissipation, and interspecies slip while avoiding wall-accommodation ambiguity.  In all DSMC tests, the Ar--Ar and He--He kernels, initial particle realization, number density, temperature, domain, and sampling protocol are held fixed; only the He--Ar kernel is changed from native EPAPS to its neural representation.

The contribution is therefore fivefold.  First, we formulate a validation ladder for neural scattering kernels that starts from pointwise angle prediction and ends with transport, measure, spectrum, and robustness diagnostics.  Second, we add a focused loss-ablation study showing that transport-aware losses can improve the learned angular map only when they are weighted moderately; the result is useful as a design and diagnostic study rather than as a monotone guarantee that every additional physics term improves training.  Third, we demonstrate that the same He--Ar neural kernel that passes these diagnostics reproduces the EPAPS composition-mode relaxation history in a DSMC binary-mixture problem, and we quantify this result over three independent particle realizations.  Fourth, we add an independent shear-wave relaxation test showing that the neural kernel also preserves the momentum-diffusion response of the mixture.  Fifth, we introduce a two-dimensional periodic shear-layer validation that moves beyond one-dimensional modal relaxation: the neural kernel reproduces the native EPAPS mixing-index history, selected-time mole-fraction field, and species-slip field in an evolving binary-mixture flow.  A Chapman--Enskog calculation from the EPAPS diffusion cross section is used to anchor the field-level apparent diffusivity to a transport-theory scale, while the primary solver-level metric remains the EPAPS-to-NN ratio so that the sheared finite-gradient field is not overinterpreted as a homogeneous CE experiment.  The framework is demonstrated for a refined argon--argon J\"ager table and a helium--argon ab initio neural surrogate based on the EPAPS data of Sharipov and Benites.  A helium--helium EPAPS table is also converted into the same equal-area format, completing the pairwise collision database needed for Ar--He mixture DSMC.

The remainder of the paper is organized to keep the physical argument compact.  In \hyperref[sec:theory]{Sec.~\ref*{sec:theory}}, we collect the theoretical and methodological ingredients: the elastic-scattering map, the equal-impact-area coordinate, the transport and representative-collision functionals, the cumulative angular measure, and the neural trigonometric representation used to approximate the He--Ar scattering map.  In \hyperref[sec:results]{Sec.~\ref*{sec:results}}, we present all validation results in a single sequence, beginning with the refined Ar--Ar J\"ager stress test in \hyperref[sec:results_arar]{Sec.~\ref*{sec:results_arar}}, followed by the main He--Ar EPAPS pair-level validation in \hyperref[sec:results_arhe]{Sec.~\ref*{sec:results_arhe}}, and then the three DSMC mixture tests: the sinusoidal composition mode in \hyperref[sec:sinusoidal_test]{Sec.~\ref*{sec:sinusoidal_test}}, the transverse shear wave in \hyperref[sec:shearwave_test]{Sec.~\ref*{sec:shearwave_test}}, and the two-dimensional periodic shear layer in \hyperref[sec:shearlayer_test]{Sec.~\ref*{sec:shearlayer_test}}.  The supporting robustness, interpolation, refinement, low-energy, and field-profile checks are placed in Appendix~\ref{app:additional_checks} so that the main text remains centered on the physical validation hierarchy.  Finally, \hyperref[sec:discussion]{Sec.~\ref*{sec:discussion}} synthesizes the implications and limitations of the results, and \hyperref[sec:conclusion]{Sec.~\ref*{sec:conclusion}} summarizes the evidence that preservation of angular measures and transport projections, rather than angle error alone, is the relevant criterion for neural ab initio scattering kernels in DSMC gas mixtures.

\section{Methods: scattering maps, transport functionals, and neural validation}\label{sec:theory}

\subsection{Elastic binary scattering}

For two particles of masses $m_i$ and $m_j$, let $\bm{g}=\bm{c}_i-\bm{c}_j$ be the relative velocity and let
\begin{equation}
  m_{ij}=\frac{m_i m_j}{m_i+m_j}
\end{equation}
be the reduced mass.  The relative collision energy is
\begin{equation}
  \Er=\frac{1}{2}m_{ij}g^2.
\end{equation}
For a central potential, the classical deflection angle can be expressed formally as
\begin{equation}
  \chi(\Er,b)=\pi-2b\int_{r_{\min}}^{\infty}
  \frac{\dd r}{r^2\left[1-b^2/r^2-\Phi(r)/\Er\right]^{1/2}},
  \label{eq:classical_scattering}
\end{equation}
where $\Phi(r)$ is the pair potential and $r_{\min}$ is the turning point.  The exact form of the integral is not the main object of this paper; rather, we assume that a high-fidelity reference map has been generated either by trajectory integration, by a refined potential-specific table, or by ab initio supplementary data.  The task is to represent and validate this map accurately enough for DSMC.

For a finite effective scattering area at fixed energy, we write
\begin{equation}
  q=\left(\frac{b}{b_{\max}(\Er)}\right)^2,\qquad 0\le q\le1.
  \label{eq:q_coordinate}
\end{equation}
Uniform sampling in $q$ is uniform sampling in impact area because
\begin{equation}
  2\pi b\,\dd b=\pi b_{\max}^2\,\dd q.
\end{equation}
Thus, an equal-area table $\chi(q_k,{\Er}_\ell)$ can be used directly to compute transport integrals by simple quadrature.  This is the natural coordinate for EPAPS-style scattering data.  For the refined Ar--Ar J\"ager table we instead use a high-resolution physical $b$ grid and perform the corresponding $2\pi b\,\dd b$ quadrature.

\subsection{Transport cross sections}

The transport cross sections are projections of the angular scattering measure.  The diffusion cross section is
\begin{equation}
  \QD(\Er)=\int (1-\cos\chi)\,\dd\sigma,
  \label{eq:QD_general}
\end{equation}
whereas the viscosity cross section is
\begin{equation}
  \Qmu(\Er)=\int (1-\cos^2\chi)\,\dd\sigma.
  \label{eq:Qmu_general}
\end{equation}
They appear in the Chapman--Enskog collision integrals and determine, respectively, diffusion-like momentum randomization and shear-momentum relaxation.  In an equal-area representation with total scattering area $\sigma_T(\Er)$ and $N_q$ points,
\begin{align}
  \QD(\Er)&\approx \frac{\sigma_T(\Er)}{N_q}\sum_{k=1}^{N_q}
    \left[1-\cos\chi(q_k,\Er)\right],\label{eq:QD_equal_area}\\
  \Qmu(\Er)&\approx \frac{\sigma_T(\Er)}{N_q}\sum_{k=1}^{N_q}
    \left[1-\cos^2\chi(q_k,\Er)\right].\label{eq:Qmu_equal_area}
\end{align}
For a continuous physical impact-parameter grid,
\begin{align}
  \QD(\Er)&=2\pi\int_0^{b_{\max}}[1-\cos\chi(\Er,b)]b\,\dd b,\label{eq:QD_b}\\
  \Qmu(\Er)&=2\pi\int_0^{b_{\max}}[1-\cos^2\chi(\Er,b)]b\,\dd b.\label{eq:Qmu_b}
\end{align}
These formulas clarify why angular regression error cannot be the only criterion.  The weights in Eqs.~\eqref{eq:QD_general} and \eqref{eq:Qmu_general} emphasize different angular regions.  Small-angle errors may be weakly visible in $\chi$ but important in aggregate if they occur over a large area; backward or side-scattering errors can strongly affect $\Qmu$ even when $\QD$ remains acceptable.  They also provide a useful physical interpretation of the curves discussed below.  The diffusion cross section is an area-weighted measure of longitudinal relative-momentum loss, since $1-\cos\chi$ is the fractional loss of forward relative momentum in a collision.  The viscosity cross section weights the decay of the second angular moment, $1-\cos^2\chi$, and is therefore more sensitive to side scattering than to purely forward or purely backward scattering.  In the small-angle limit, $1-\cos\chi\simeq\chi^2/2$ whereas $1-\cos^2\chi\simeq\chi^2$, so $\Qmu/\QD$ tends toward 2 for a purely forward-focused map.  When attractive interactions or orbiting/rainbow-like branches produce large-angle scattering, this ratio decreases and may vary nonmonotonically with energy.  The energy dependence of $\QD$ and $\Qmu$ therefore encodes a competition between the long interaction time and large effective capture area at low collision energy, the rapid passage and smaller deflection area at high energy, and intermediate-energy branch structure where the deflection function changes rapidly with impact parameter.

\subsection{Ohr-style representative quantities}

Ohr introduced a representative single-deflection formulation designed to reproduce the two transport cross sections without relying on an arbitrary total cross section \cite{Ohr2023}.  The representative collision cross section is
\begin{equation}
  \RCS(\Er)=\frac{\QD(\Er)^2}{2\QD(\Er)-\Qmu(\Er)},
  \label{eq:RCS}
\end{equation}
and the representative deflection angle is defined by
\begin{equation}
  \cos\RDA(\Er)=\frac{\Qmu(\Er)}{\QD(\Er)}-1.
  \label{eq:RDA}
\end{equation}
The equivalent VSS parameter and cross section are
\begin{align}
  \alpha_{\rm VSS}(\Er)&=\frac{2\Qmu(\Er)}{2\QD(\Er)-\Qmu(\Er)},\label{eq:alpha_vss} \\
  \SigVSS(\Er)&=\frac{\QD(\Er)}{2}\frac{2\QD(\Er)+\Qmu(\Er)}{2\QD(\Er)-\Qmu(\Er)}.\label{eq:sigma_vss}
\end{align}
These combinations are nonlinear.  A surrogate may preserve $\QD$ and $\Qmu$ separately within small error yet amplify errors in $\RCS$ or $\alpha_{\rm VSS}$ if the denominator $2\QD-\Qmu$ is sensitive.  Physically, $\RCS$ is the collision area that a single representative deflection would need in order to reproduce the simultaneous diffusion and viscosity projections of the full angular distribution.  The representative angle is not an arbitrary fit parameter: through $\cos\RDA=\Qmu/\QD-1$, it measures how much of the same momentum-randomizing area is associated with side-scattering versus forward/backward scattering.  When $\cos\RDA$ is small or negative, the equivalent single-angle picture represents a broad or side-weighted angular redistribution; when it approaches unity, the scattering is strongly forward-focused.  We therefore include Eqs.~\eqref{eq:RCS}--\eqref{eq:sigma_vss} as independent validation targets and as compact physical diagnostics of the angular redistribution.

\subsection{Cumulative angular measure}

The angular push-forward measure is defined by mapping impact area into $\mu=\cos\chi$.  The normalized cumulative angular measure is
\begin{equation}
  \Sigma(\mu;\Er)=\frac{1}{\sigma_T(\Er)}\int_{\cos\chi\le \mu}\dd\sigma.
  \label{eq:cumulative}
\end{equation}
For an equal-area table, $\Sigma$ is simply the cumulative distribution function of the set $\{\cos\chi(q_k,\Er)\}_{k=1}^{N_q}$.  This diagnostic is deliberately derivative-free.  The differential cross section $\dd\sigma/\dd\Omega$ requires differentiating a possibly multi-valued or turning-point-containing scattering map, and it can become singular where $\dd\chi/\dd b\to0$.  By contrast, $\Sigma(\mu)$ is well defined even when the branch structure is complicated.  The shape of $\Sigma$ has a direct physical interpretation.  A curve that rises only near $\mu=1$ indicates that most impact area produces small-angle, forward scattering.  A broad rise over $-1<\mu<1$ indicates substantial side and backward scattering, which is precisely the part of the distribution that affects diffusion and viscosity cross sections differently.  Sharp shoulders or steps signal impact-parameter intervals that map into a narrow angular range, the cumulative counterpart of rainbow or turning behavior in the differential picture.  Agreement in $\Sigma$ therefore implies that the surrogate has preserved the angular redistribution of impact area, not only a pointwise curve.

\subsection{Neural representation and loss-informed diagnostics}\label{sec:surrogate}

\subsubsection{Input and output representation}

The neural scattering surrogate approximates
\begin{equation}
  \mathcal{G}_\theta:(\log \Er,q)\mapsto (\cos\chi,\sin\chi).
  \label{eq:network_map}
\end{equation}
The trigonometric output is used because direct regression on $\chi$ can introduce artificial discontinuities near angular wrapping points.  The angle is reconstructed as
\begin{equation}
  \chi_\theta=\operatorname{atan2}\left(s_\theta,c_\theta\right),
\end{equation}
with appropriate mapping to the physical interval used by the reference data.  Training and validation are performed in $\log\Er$ because the scattering map varies over many decades in collision energy.  For EPAPS equal-area tables, $q$ is the natural second input.  For physical $b$-grid data, one may use $b/b_{\max}$ or $q=(b/b_{\max})^2$ depending on the quadrature and sampling objective.

The Ar--Ar Lennard--Jones/J\"ager surrogate used in our earlier work was a DeepONet representation~\cite{Lu2021DeepONet}.  A branch network encoded the reduced energy and a trunk network encoded the impact coordinate; their latent features were combined by an inner product.  The final dataset contained $1.28\times10^7$ valid scattering samples, split into $1.152\times10^7$ training samples and $1.28\times10^6$ validation samples, with $10^{-3}\le\eps^\ast\le40$ and $0\le b/b_{\max}\le1$.  The DeepONet used a latent dimension of 128, branch and trunk subnetworks of width 128 with four hidden layers, Adam optimization, learning rate $10^{-3}$, and mini-batches of 8192.  Held-out validation in that work gave a bulk mean wrapped-angle error of order $10^{-3}$ rad and high-percentile errors below $10^{-2}$ rad.

For the He--Ar EPAPS data used here, the native table contains 900 collision-energy levels and 100 equal-impact-area samples per energy.  The production neural surrogate is a fully connected trigonometric-output network trained on the map $(\log \Er,q)\to(\cos\chi,\sin\chi)$ over $\Er/\kb\simeq4.9\times10^{-4}$--$1.53\times10^5~\mathrm{K}$.  The network uses five hidden layers of width 192 with smooth nonlinear activation, batch size 8192, 2500 epochs, and low-energy sample weighting to prevent the high-density high-energy region from dominating the fit.  A transport-aware penalty is evaluated periodically during training on selected energy batches.  The final network output is exported back into a DSMC-ready equal-area table, so that the online collision kernel can use fast table lookup or neural inference depending on the implementation.

\paragraph*{Reproducibility protocol.}
For the He--Ar production network, the input features are the affine-scaled variables $\tilde E=2(\log E_r-\log E_{\min})/(\log E_{\max}-\log E_{\min})-1$ and $\tilde q=2q-1$, and the two network outputs are renormalized to the unit circle before reconstructing $\chi_\theta$.  The training/validation split is performed in the flattened $(E_r,q)$ sample set but is checked by energy-resolved diagnostics rather than by a single random MSE alone.  The production and ablation runs used the same five-hidden-layer, width-192 architecture, Adam optimization, mini-batches of 8192, and a fixed random seed recorded in the source metadata.  The low-energy emphasis is implemented through log-energy-aware sampling and weighting so that the $1$--$100~\mathrm{K}$ interval is not overwhelmed by the larger high-energy portion of the table.  The transport loss is evaluated on energy batches using the equal-area quadrature in Eqs.~\eqref{eq:QD_equal_area}--\eqref{eq:Qmu_equal_area}; the cumulative-measure and spectral diagnostics are evaluated only as validation quantities for the production surrogate.  The exact training script, exported table, random seeds, training log, and plotting scripts are part of the review-access source package, so the numerical values in Tables~\ref{tab:loss_ablation} and \ref{tab:summary} can be regenerated from the same inputs.

\subsubsection{Pointwise, transport, and measure losses}

The basic angular loss is
\begin{equation}
  \mathcal{L}_{\chi}=\frac{1}{N_b}\sum_{n=1}^{N_b}
  \left[(c_{\theta,n}-c_{n}^{\rm ref})^2+(s_{\theta,n}-s_{n}^{\rm ref})^2\right],
  \label{eq:Lchi}
\end{equation}
where $c=\cos\chi$ and $s=\sin\chi$.  A transport-augmented objective adds batch estimates of $\QD$ and $\Qmu$:
\begin{equation}
\begin{aligned}
  \mathcal{L}_{Q}
  &=\frac{1}{N_E}\sum_{\ell=1}^{N_E}
  \bigg[\left(\frac{Q_{D,\theta}(E_{r,\ell})}
  {Q_{D,\rm ref}(E_{r,\ell})}-1\right)^2 \\
  &\hspace{1.1cm}+\left(\frac{Q_{\mu,\theta}(E_{r,\ell})}
  {Q_{\mu,\rm ref}(E_{r,\ell})}-1\right)^2\bigg].
\end{aligned}
  \label{eq:LQ}
\end{equation}
A measure-aware objective can further include
\begin{equation}
\begin{aligned}
  \mathcal{L}_{\Sigma}
  &=\frac{1}{N_E}\sum_{\ell=1}^{N_E}
  \left\|\Sigma_\theta(\mu;E_{r,\ell})
  -\Sigma_{\rm ref}(\mu;E_{r,\ell})\right\|_2^2 .
\end{aligned}
  \label{eq:LSigma}
\end{equation}
To test whether these diagnostics are only post-processing quantities or can also guide the learning problem, we performed a compact loss-ablation study.  Five He--Ar networks with the same architecture, data split, low-energy weighting, and optimization schedule were trained using
\begin{equation}
  \mathcal{L}=\mathcal{L}_{\chi}+\lambda_Q\mathcal{L}_Q+\lambda_\Sigma\mathcal{L}_\Sigma+\lambda_s\mathcal{L}_{\rm smooth},
  \label{eq:combined_loss}
\end{equation}
where the small smoothness term was kept fixed when present in the production training script.  The ablation included angle-only training, two transport-aware cases with $\lambda_Q=0.05$ and $0.10$, and two combined transport/measure cases with $(\lambda_Q,\lambda_\Sigma)=(0.05,0.01)$ and $(0.10,0.02)$.  Table~\ref{tab:loss_ablation} summarizes the validation behavior.

\begin{table*}[t]
\centering
\small
\caption{Loss-ablation study for the He--Ar neural scattering surrogate.  All entries use the same network architecture and data split.  The validation number is the mean $\mathcal{L}_{\chi}$ over the final 50 epochs; the improvement is measured relative to the angle-only baseline.  A modest transport-aware penalty gives the best angular generalization, while larger transport weight or adding a cumulative-measure penalty introduces an optimization trade-off in the present training setup.}
\label{tab:loss_ablation}
\begin{tabular}{lcccc}
\toprule
Training objective & $\lambda_Q$ & $\lambda_\Sigma$ & Final-window $\mathcal{L}_{\chi}^{\rm val}$ & Change vs. angle-only \\
\midrule
Angle only & 0 & 0 & $6.71\times10^{-4}$ & reference \\
Transport-aware & 0.05 & 0 & $5.75\times10^{-4}$ & $14.4\%$ lower \\
Transport-aware & 0.10 & 0 & $7.02\times10^{-4}$ & $4.7\%$ higher \\
Transport + measure & 0.05 & 0.01 & $7.01\times10^{-4}$ & $4.4\%$ higher \\
Transport + measure & 0.10 & 0.02 & $7.57\times10^{-4}$ & $12.8\%$ higher \\
\bottomrule
\end{tabular}
\end{table*}

The ablation supports two conclusions that are important for interpreting the rest of the paper.  First, transport information is not only a diagnostic: a modest $\mathcal{L}_Q$ contribution improves the learned angular map, consistent with the idea that the network should see physically relevant projections during training.  Second, the effect is not monotone in loss complexity.  Increasing $\lambda_Q$ or adding the cumulative-measure penalty with the present weights improves neither the final validation angle loss nor the stability of the training curves.  Therefore, the production surrogate is best described as a tuned transport-aware representation, not as a generic result of adding arbitrary physics penalties.

\subsubsection{Spectral content of the scattering map}

To connect with spectrally informed flow reconstruction, we define a Fourier-amplitude diagnostic in the equal-area coordinate $q$.  For each energy level, the $q$-mean is subtracted:
\begin{equation}
  \chi'(q,\Er)=\chi(q,\Er)-\langle \chi(q,\Er)\rangle_q.
\end{equation}
The discrete Fourier coefficients are
\begin{equation}
  \widehat{\chi}_k(\Er)=\sum_{m=0}^{N_q-1}\chi'(q_m,\Er)
  \exp\left(-\frac{2\pi i k m}{N_q}\right),
  \label{eq:fft}
\end{equation}
with spectral energy
\begin{equation}
  S_k(\Er)=|\widehat{\chi}_k(\Er)|^2.
  \label{eq:spectral_energy}
\end{equation}
The spectral-amplitude discrepancy is then measured by
\begin{equation}
  \mathcal{E}_{\rm spec}(\Er)=\left[\frac{\sum_k\left(\log S_{k,\theta}-\log S_{k,\rm ref}\right)^2}{\sum_k\left(\log S_{k,\rm ref}\right)^2}\right]^{1/2},
  \label{eq:Espec}
\end{equation}
with a small floor added in practice before the logarithm.  We also compute a high-mode energy fraction
\begin{equation}
  F_{\rm high}(\Er)=\frac{\sum_{k\ge k_c}S_k(\Er)}{\sum_{k\ge1}S_k(\Er)},
  \label{eq:high_mode_fraction}
\end{equation}
and report $F_{\rm high}^{\NN}/F_{\rm high}^{\EPAPS}$.  This diagnostic asks whether the surrogate preserves sharp or oscillatory features in $q$, not only low-order angular trends.

\section{Results}\label{sec:results}

\subsection{Refined Ar--Ar J\"ager validation}\label{sec:results_arar}

The first test is a refined Ar--Ar J\"ager scattering table.  The low-energy J\"ager map contains turning and orbiting features that are difficult to represent with a coarse impact-parameter grid.  We therefore regenerated a refined table over $\Er/\kb=40$--$10500~\mathrm{K}$ with 1600 impact-parameter points and adaptive refinement near high-curvature regions.  The resulting table is used here as the Ar--Ar component of a future Ar--He mixture database and as a reference example of why pointwise angular and transport validation should be separated.

\begin{figure*}[t]
\centering
\includegraphics[width=0.96\textwidth,trim=0 0 0 0.42in,clip]{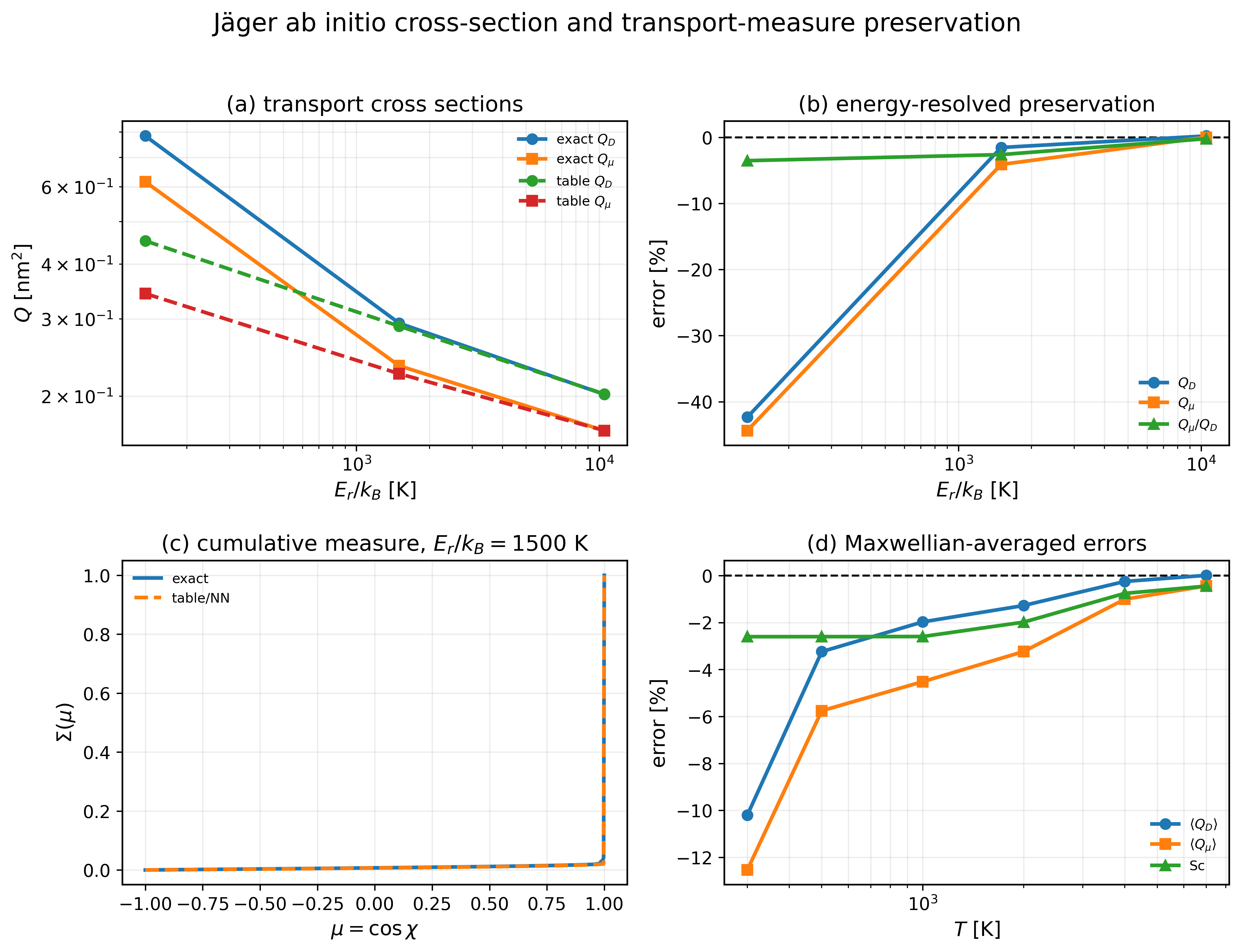}
\caption{Refined Ar--Ar J\"ager validation dashboard.  The panels combine the deflection map, transport cross sections, transport-preservation errors, and cumulative angular measure.  The refined table resolves low-energy structure while preserving $\QD$, $\Qmu$, and angular-measure statistics.}
\label{fig:arar_dashboard}
\end{figure*}

Figure~\ref{fig:arar_dashboard} summarizes the refined Ar--Ar result.  The deflection map resolves the low-energy branch structure; the transport cross sections decrease with increasing energy because high-energy pairs spend less time in the attractive well and only a smaller impact-parameter interval produces appreciable angular deflection.  At lower energy, by contrast, the attractive tail and the centrifugal barrier allow larger impact areas to contribute to momentum randomization, which increases both $\QD$ and $\Qmu$.  The remaining nonmonotone features are not numerical noise; they are the transport-integral signature of rapid changes in the deflection function near turning/orbiting branches.  The viscosity projection responds somewhat differently from the diffusion projection because side-scattering weights $1-\cos^2\chi$ more strongly than $1-\cos\chi$.  The cumulative angular measure provides a derivative-free confirmation that the impact-area distribution in angle space is preserved.  We do not use the differential cross section $\dd\sigma/\dd\Omega$ as a primary validation target because it requires differentiating a multi-branched map and becomes singular near turning points.  The cumulative measure is more stable and more directly linked to DSMC angular sampling.

\subsection{He--Ar EPAPS neural scattering surrogate}\label{sec:results_arhe}

\subsubsection{Deflection maps and transport cross sections}

\begin{figure*}[t]
\centering
\includegraphics[width=0.96\textwidth,trim=0 0 0 0.28in,clip]{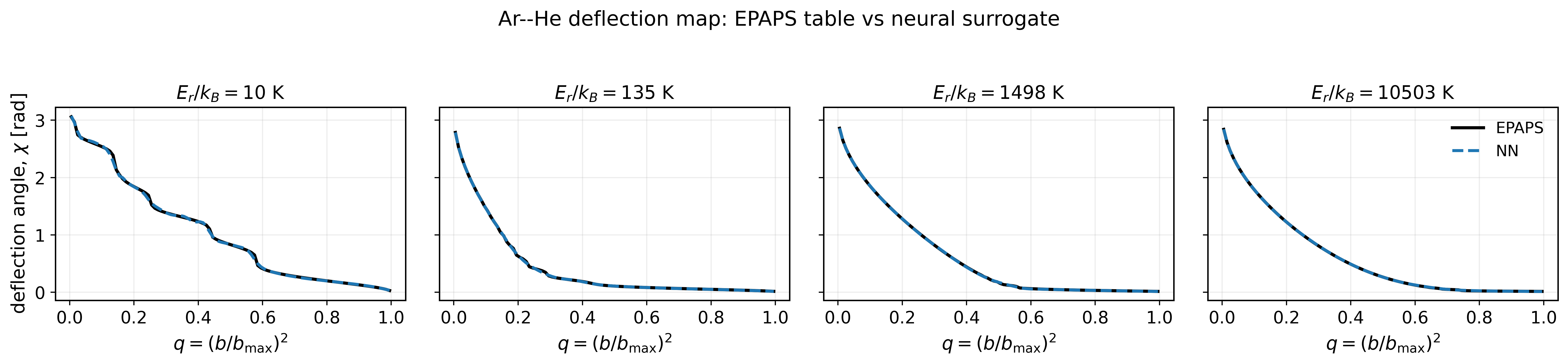}
\includegraphics[width=0.96\textwidth,trim=0 0 0 0.32in,clip]{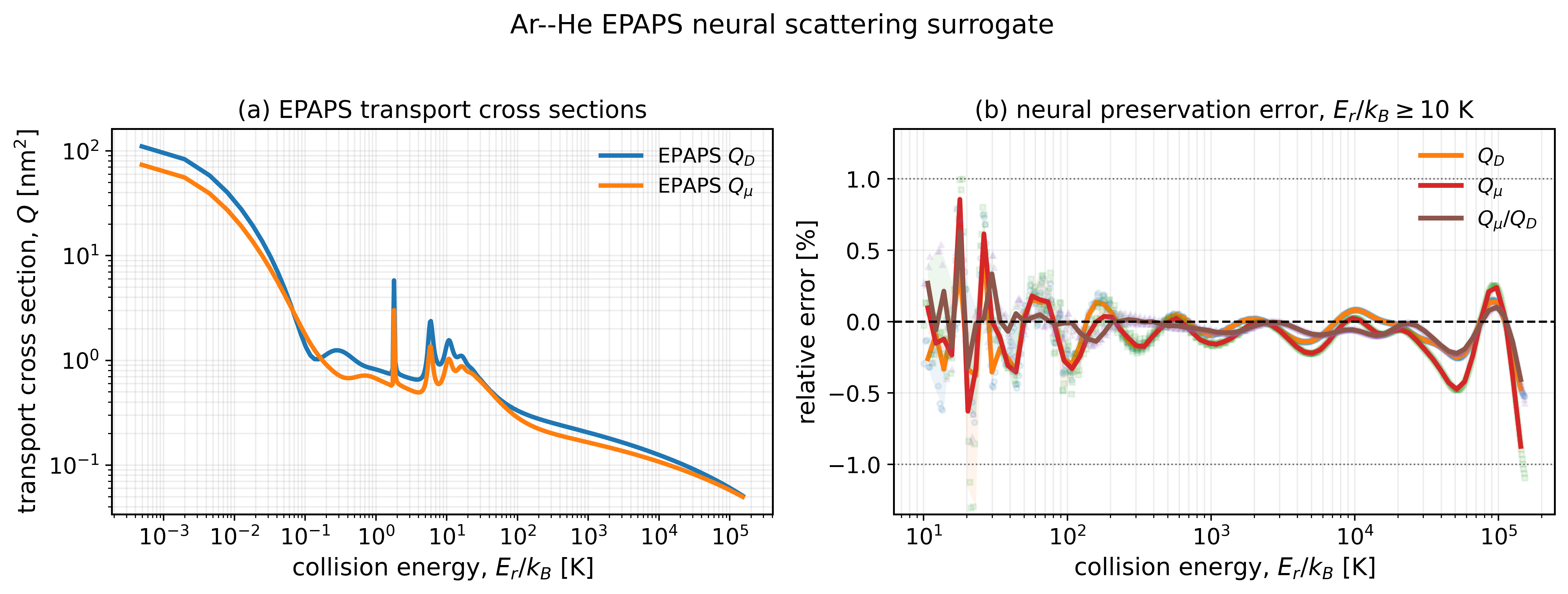}
\caption{He--Ar ab initio scattering surrogate.  Top: deflection maps at selected collision energies, using the equal-impact-area coordinate $q=(b/b_{\max})^2$.  Bottom: EPAPS transport cross sections and neural preservation errors for $\QD$, $\Qmu$, and $\Qmu/\QD$.}
\label{fig:arhe_deflection_transport}
\end{figure*}

The He--Ar reference data are based on the ab initio EPAPS table of Sharipov and Benites.  The native table contains 900 energy levels and 100 equal-area samples.  The neural surrogate is exported on the same coordinate, which allows one-to-one comparison of $\chi(q,E)$ and direct evaluation of Eqs.~\eqref{eq:QD_equal_area} and \eqref{eq:Qmu_equal_area}.  Figure~\ref{fig:arhe_deflection_transport} shows the deflection map at four representative energies and the transport cross-section preservation.  At $\Er/\kb=10~\mathrm{K}$, the map contains multiple low-energy features.  By $135~\mathrm{K}$ the map is smoother but still nontrivial; at $1498$ and $10503~\mathrm{K}$, the scattering becomes increasingly forward-focused.  The neural surrogate tracks these regimes without visible loss of structure.

The transport panel gives the more important test, but it also contains useful physics.  The EPAPS $\QD$ and $\Qmu$ curves are largest at very low relative energy because slow He--Ar pairs are deflected over a large effective impact area by the attractive part of the potential.  As $\Er$ increases, the collision time shortens and forward scattering dominates over most of the impact-area interval, so both transport cross sections decay.  The decay is not a featureless power law.  The small peaks and shoulders in the $1$--$20~\mathrm{K}$ range are the transport-integral footprint of rapid changes in $\chi(q,E)$ caused by the competition between the attractive well and the repulsive core.  In this region, small shifts of a branch in $q$ can alter a narrow angular interval without strongly changing the global map.  At high energy, the curves become smoother because the scattering is increasingly controlled by short-range repulsion and most collisions are forward-focused.  For $\Er/\kb\ge10~\mathrm{K}$, the maximum neural deviations are $0.75\%$ in $\QD$, $1.37\%$ in $\Qmu$, and $0.84\%$ in $\Qmu/\QD$.  The slightly higher error in $\Qmu$ is physically expected because the viscosity weighting is more sensitive to side-scattering and angular redistribution.  The result shows that the network is not merely visually reproducing $\chi(q)$; it is preserving the transport projections that control binary diffusion and momentum relaxation in DSMC mixture transport.

\begin{figure*}[t]
\centering
\includegraphics[width=0.98\textwidth]{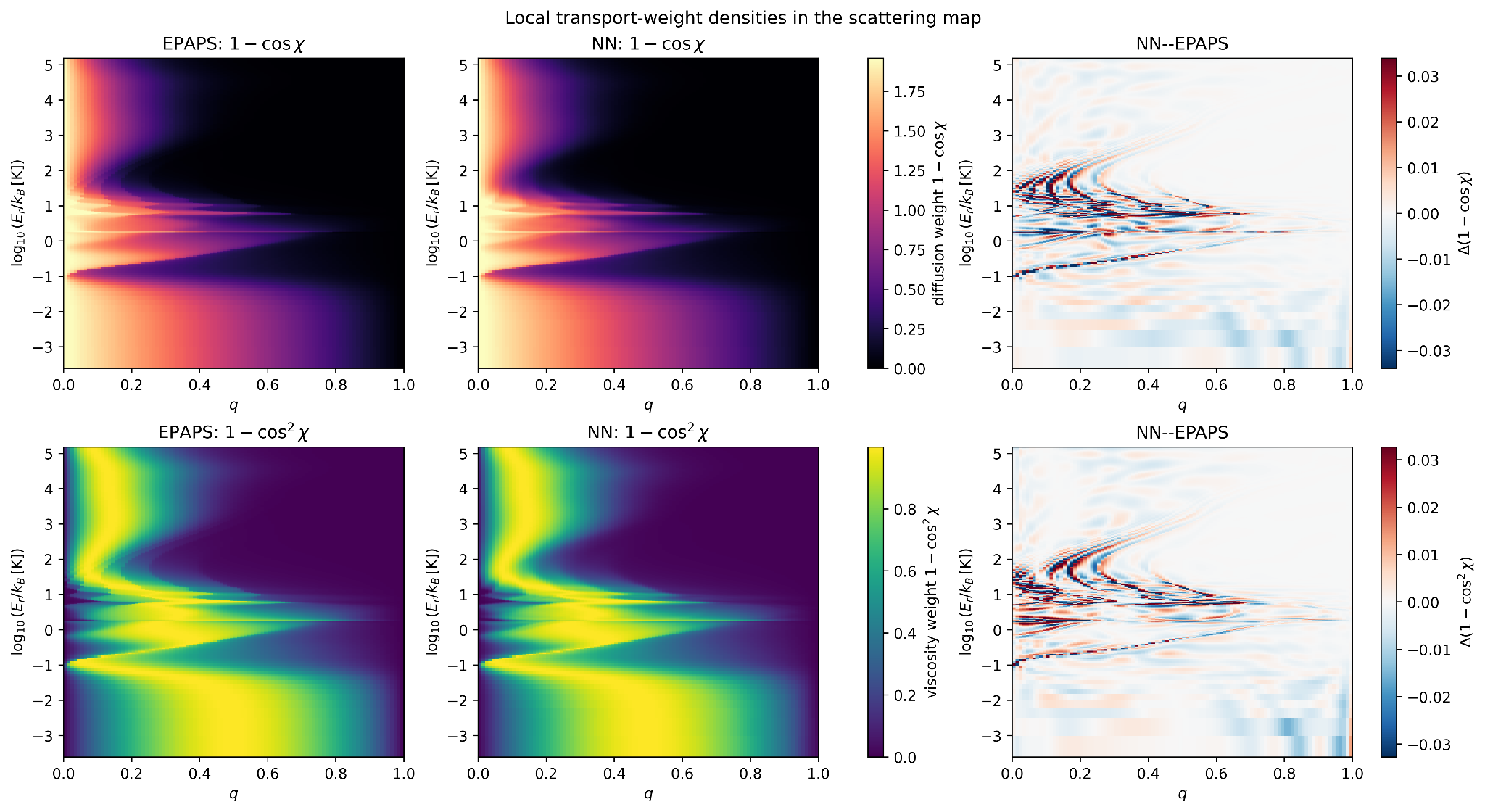}
\caption{Impact-area-resolved transport-weight densities for the He--Ar scattering map.  The upper row shows the diffusion weight $1-\cos\chi$ and its NN--EPAPS difference; the lower row shows the viscosity weight $1-\cos^2\chi$ and its difference.  These maps identify which regions of the $(E_r,q)$ plane actually contribute to $Q_D$ and $Q_\mu$.  The structured ridges arise from rapid changes of the deflection angle with impact area in the low- and intermediate-energy regimes, while the high-energy forward-scattering region carries little weight and little surrogate discrepancy.}
\label{fig:transport_weight_density_qE}
\end{figure*}

Figure~\ref{fig:transport_weight_density_qE} gives a more local physical interpretation of the same transport projections.  The diffusion weight density, $1-\cos\chi$, is large wherever an accepted collision produces a finite loss of relative-velocity direction.  It is therefore strongest in the low-energy and moderate-impact-area bands where the attractive well bends trajectories strongly, and it decays toward the high-energy/large-$q$ forward-scattering region.  The viscosity weight density, $1-\cos^2\chi$, emphasizes transverse redirection of the relative velocity; equivalently, it suppresses events close to $\cos\chi=\pm1$ and is largest for side-scattering collisions.  Consequently, the $Q_\mu$ map selects a narrower and more structured subset of the $(E_r,q)$ plane than the $Q_D$ map.  This explains why $Q_\mu$ can be more sensitive than $Q_D$ even when the two cross sections have similar overall trends.  The ridge-like structures in both weight maps are the local, impact-area-resolved origin of the oscillations and shoulders seen in the line plots: they mark energies for which a small change in impact area moves a trajectory between weak deflection, side scattering, and stronger angular redirection.  The corresponding NN--EPAPS panels show that the largest local differences are concentrated around these branch-sensitive ridges, while the high-energy forward-scattering region is nearly featureless.  Thus Fig.~\ref{fig:transport_weight_density_qE} clarifies that the physically important question is not simply where $\chi$ is large, but where surrogate error overlaps the transport weights that feed the collision integrals.

\subsubsection{Cumulative angular measure}

\begin{figure*}[t]
\centering
\includegraphics[width=0.96\textwidth,trim=0 0 0 0.38in,clip]{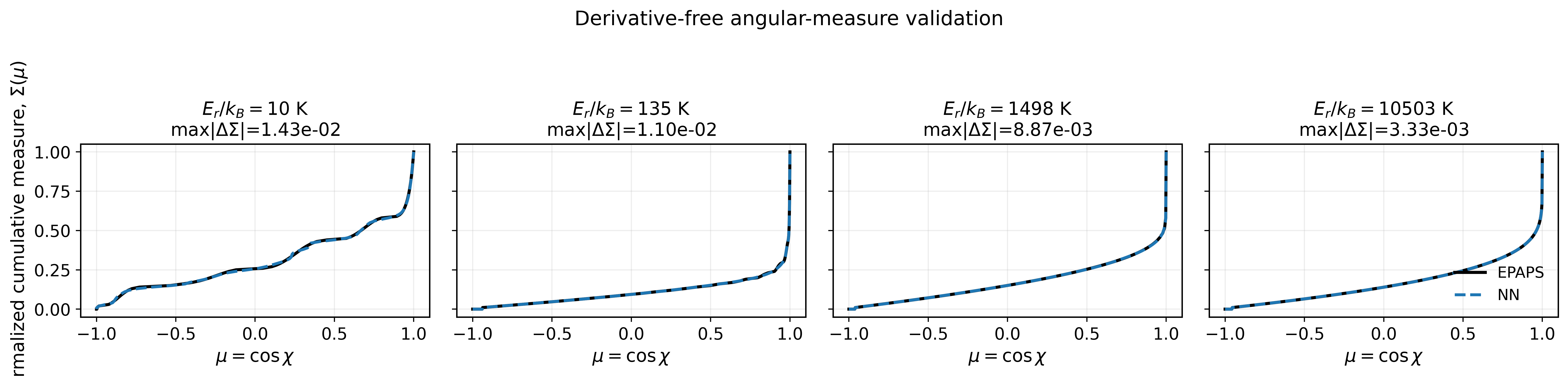}
\caption{Derivative-free angular-measure validation for He--Ar.  The cumulative angular measure $\Sigma(\mu)$ is computed from the impact-area push-forward.  Agreement between EPAPS and the neural surrogate confirms that the surrogate preserves angular scattering statistics beyond pointwise angle error.}
\label{fig:arhe_cumulative}
\end{figure*}

Figure~\ref{fig:arhe_cumulative} compares $\Sigma(\mu)$ for the same four energies.  The cumulative measure is particularly useful at low energy because it does not require branch identification or numerical differentiation.  Its physical trend mirrors the cross-section behavior.  At lower energy, $\Sigma(\mu)$ rises over a wider range of $\mu$, indicating that a non-negligible fraction of impact area produces side or backward scattering.  At higher energy, the rise is compressed toward $\mu=1$, showing that most accepted collisions are small-angle and forward-focused.  The shoulders in the low-energy cumulative curves are the measure-space signature of impact-parameter intervals that are focused into similar deflection angles; this is the robust cumulative analogue of rainbow-like structure in a differential cross section.  Agreement in $\Sigma$ means that the area fraction of collisions producing scattering below a given $\mu$ is correct.  For DSMC, this is the relevant stochastic object: it determines how impact-area samples populate angular outcomes.  The maximum deviations in the normalized cumulative measure are about $1.4\times10^{-2}$ at $10~\mathrm{K}$ and decrease at higher energy.

\begin{figure*}[t]
\centering
\includegraphics[width=0.98\textwidth]{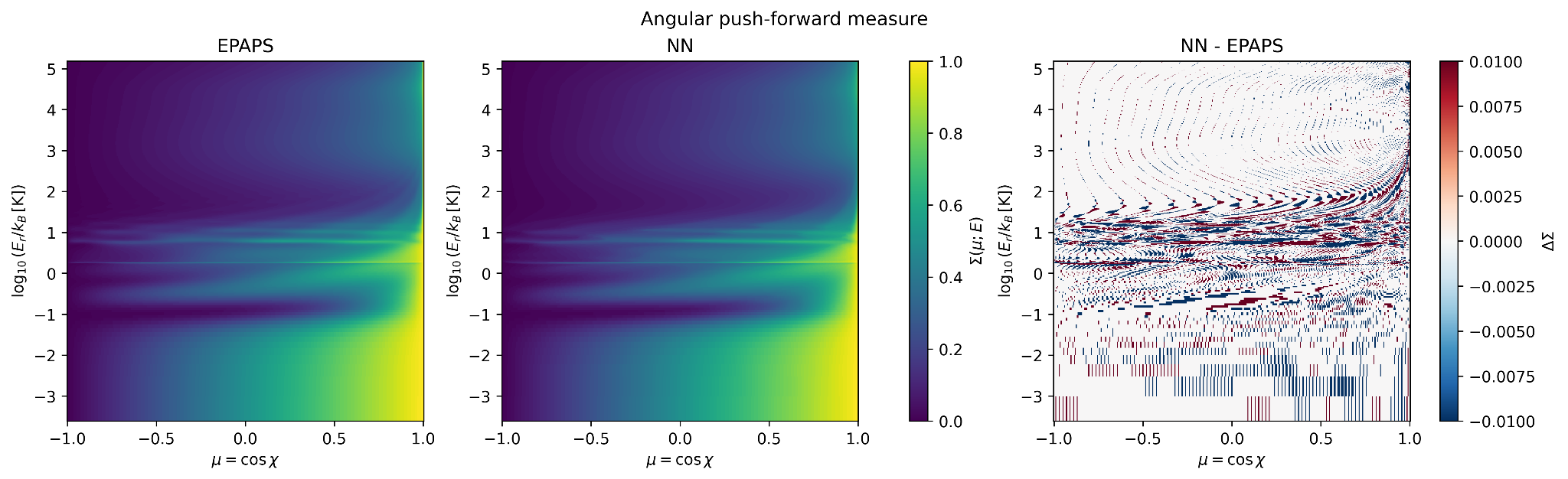}
\caption{Global angular push-forward measure for the He--Ar EPAPS and neural scattering maps.  The surface $\Sigma(\mu,E)$ gives the fraction of equal impact area that produces $\cos\chi\le\mu$.  High-energy collisions are concentrated near $\mu=1$ because scattering is predominantly forward focused, whereas low- and intermediate-energy collisions populate a broader angular range.  The difference surface is localized mainly around branch-sensitive low-energy bands, confirming that the neural kernel preserves the angular redistribution measure used by DSMC sampling.}
\label{fig:cumulative_measure_surface}
\end{figure*}

The four one-dimensional cumulative curves are useful checkpoints, but the full push-forward measure is a two-dimensional surface over $(\mu,E_r)$.  Figure~\ref{fig:cumulative_measure_surface} shows this surface directly.  At high relative energy the measure is compressed near $\mu\simeq1$, reflecting the dominance of forward scattering over almost the entire impact-area interval.  At lower energy the surface broadens toward smaller $\mu$, meaning that side-scattering and partial backscattering occupy a larger fraction of impact area.  The horizontal banding and sharp gradients in the low- and intermediate-energy region are the measure-level imprint of focusing and turning structure in the underlying potential: several nearby impact areas can be mapped into similar angular outcomes, producing shoulders in $\Sigma(\mu,E)$.  This representation is more physical than a differential cross section for the present purpose because it remains finite and interpretable even when the angular map has branch-sensitive or nearly singular regions.  The NN--EPAPS difference panel shows that the neural surrogate preserves the global push-forward surface to within small, localized deviations; therefore the stochastic angular redistribution used by DSMC is preserved over the entire EPAPS energy range, not only at the four energies shown in Fig.~\ref{fig:arhe_cumulative}.

\subsubsection{Ohr-style representative collision quantities}

\begin{figure*}[t]
\centering
\includegraphics[width=0.96\textwidth,trim=0 0 0 0.45in,clip]{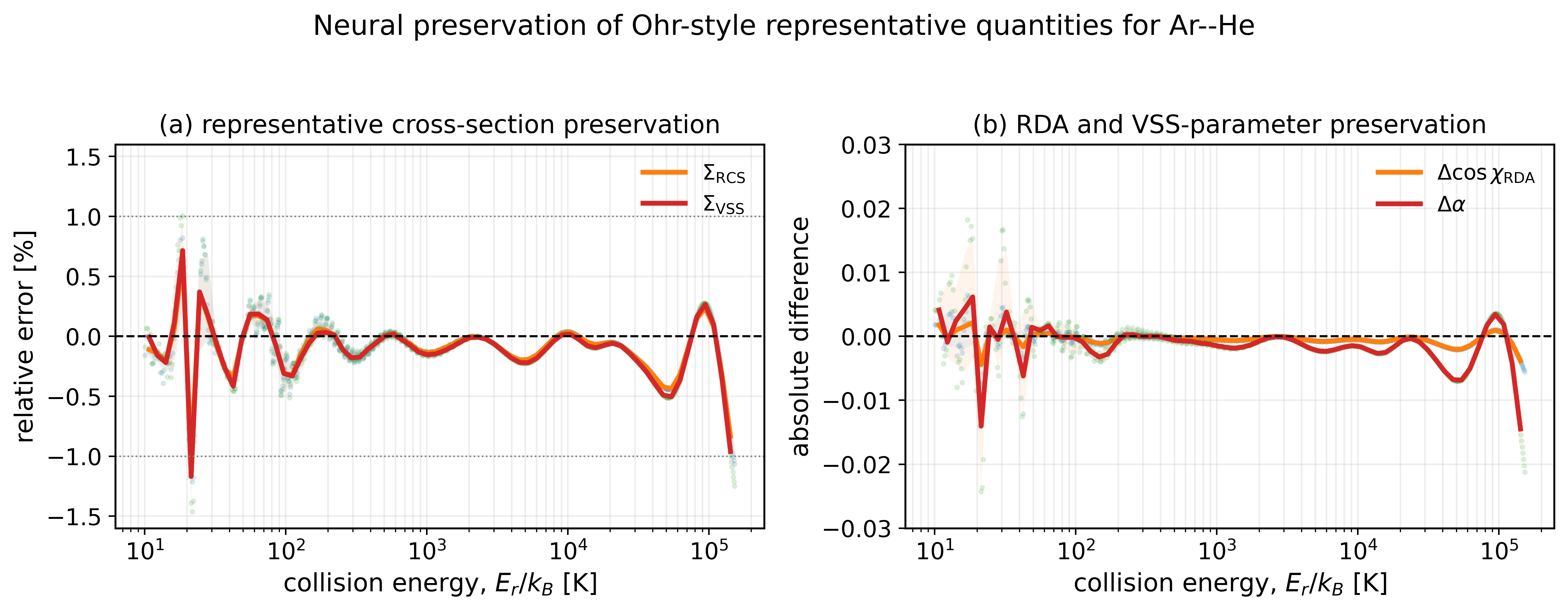}
\caption{Ohr-style neural preservation for He--Ar.  The neural surrogate preserves representative cross sections $\RCS$ and $\SigVSS$ and the corresponding representative-deflection and equivalent-VSS parameters across the DSMC-relevant energy range.}
\label{fig:ohr}
\end{figure*}

Figure~\ref{fig:ohr} tests the representative quantities defined by Eqs.~\eqref{eq:RCS}--\eqref{eq:sigma_vss}.  These quantities are important because they connect the full scattering map to reduced DSMC collision models.  They also reveal how the angular distribution changes with energy.  When the representative deflection diagnostic $\cos\RDA=\Qmu/\QD-1$ is close to unity, the transport response can be represented by a small effective deflection acting over a comparatively large area; when it is smaller or negative, side-scattering and broad angular redistribution are more important.  The oscillations in $\cos\RDA$ and in the representative cross sections occur where $\QD$ and $\Qmu$ respond differently to the same branch structure in $\chi(q,E)$.  This explains why a reduced single-deflection model must be constructed from both cross sections rather than from a single total collision area.  If the neural surrogate distorted $\QD$ and $\Qmu$ in a correlated way, the error might not be apparent from either cross section alone but would appear in $\RCS$, $\SigVSS$, $\cos\RDA$, or $\alpha_{\rm VSS}$.  The maximum errors over $\Er/\kb\ge10~\mathrm{K}$ are $1.21\%$ for $\RCS$ and $1.46\%$ for $\SigVSS$.  The maximum absolute deviations in $\cos\RDA$ and $\alpha_{\rm VSS}$ are $7.5\times10^{-3}$ and $2.4\times10^{-2}$, respectively.  This provides an independent check that the surrogate preserves representative collision statistics, not only raw transport integrals.

\subsubsection{Spectral preservation of the scattering map}

\begin{figure*}[t]
\centering
\includegraphics[width=0.96\textwidth,trim=0 0 0 0.45in,clip]{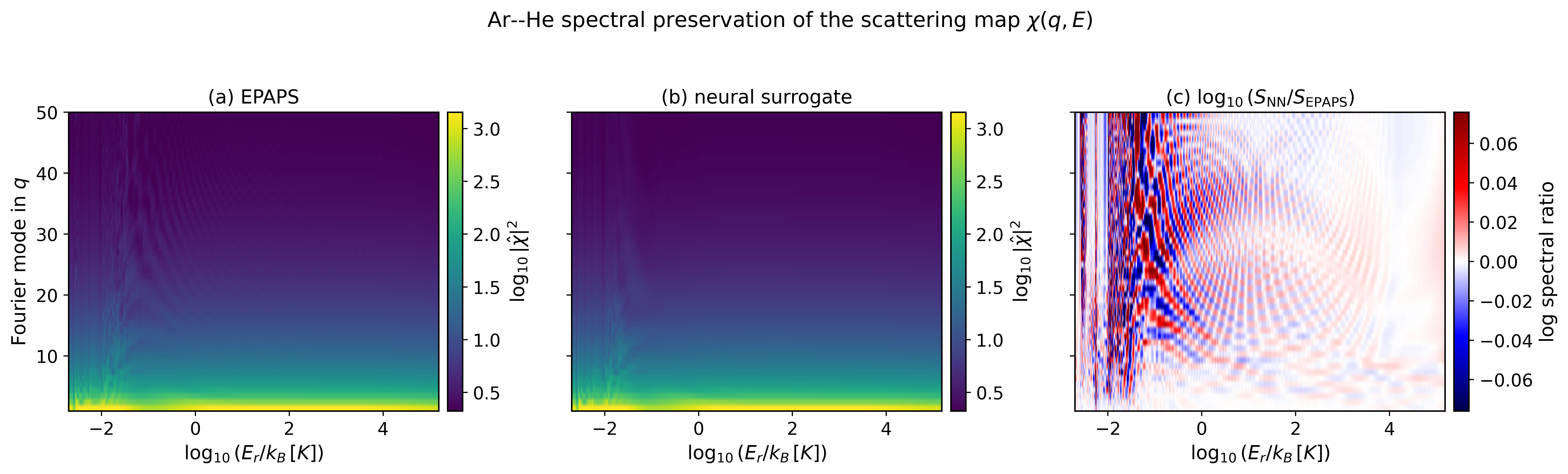}
\caption{Spectral preservation of the He--Ar scattering map $\chi(q,E)$.  The Fourier spectrum is computed in the equal-area coordinate $q$ after subtracting the $q$-mean.  The neural surrogate reproduces the EPAPS spectral content over the full energy range, with visible deviations concentrated in the low-energy, high-mode region.}
\label{fig:spectral}
\end{figure*}

Figure~\ref{fig:spectral} compares the Fourier-mode energy of the EPAPS and neural scattering maps.  The low modes dominate, as expected for a smooth impact-area map, but high modes carry the sharp low-energy features.  This separation has a direct physical meaning.  Low modes represent the slowly varying deflection produced by the overall attractive and repulsive length scales of the potential; high modes represent narrow impact-parameter intervals where the trajectory is rapidly redirected by a turning point, orbiting-like passage, or branch transition.  At high collision energy these sharp structures weaken because the map is dominated by forward scattering and short interaction time.  At low energy, the attractive well stretches the trajectory and amplifies high-curvature features in $q$.  The neural surrogate reproduces the spectral energy over the full range of $\log_{10}(\Er/\kb)$; the logarithmic ratio shows that the largest deviations occur in the low-energy/high-mode region, precisely where the scattering physics is most branch-sensitive.

\begin{figure*}[t]
\centering
\includegraphics[width=0.90\textwidth,trim=0 0 0 0.45in,clip]{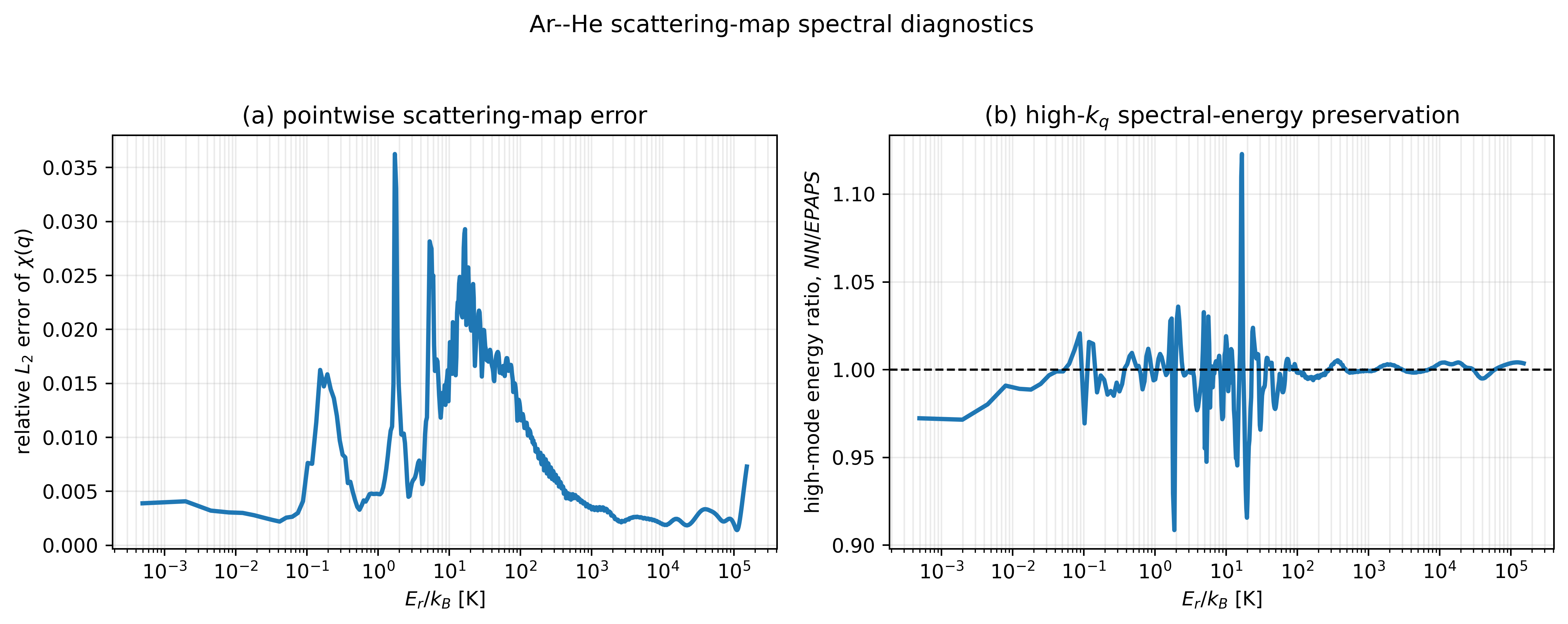}
\caption{Compact spectral diagnostics for He--Ar.  The pointwise scattering-map error is largest in the difficult $1$--$100~\mathrm{K}$ range, while the high-mode spectral-energy ratio remains close to unity over most of the EPAPS domain.}
\label{fig:spectral_diag}
\end{figure*}

Figure~\ref{fig:spectral_diag} condenses the spectral comparison.  The median relative $L_2$ error of $\chi(q)$ is $3.4\times10^{-3}$; the maximum is $3.6\times10^{-2}$.  The high-mode energy ratio has median $1.00008$, indicating essentially unbiased high-mode preservation in aggregate.  These results support the use of spectral diagnostics as a complement to transport cross sections.  They also provide a bridge to loss-function design: the ablation in Sec.~\ref{sec:surrogate} shows that transport-aware penalties can improve generalization, and an explicitly spectral term based on Eq.~\eqref{eq:Espec} is a natural next extension if one wants to target low-energy high-mode errors more aggressively.

\subsubsection{Coarse-impact and angular-noise robustness}

\begin{figure*}[t]
\centering
\includegraphics[width=0.96\textwidth,trim=0 0 0 0.45in,clip]{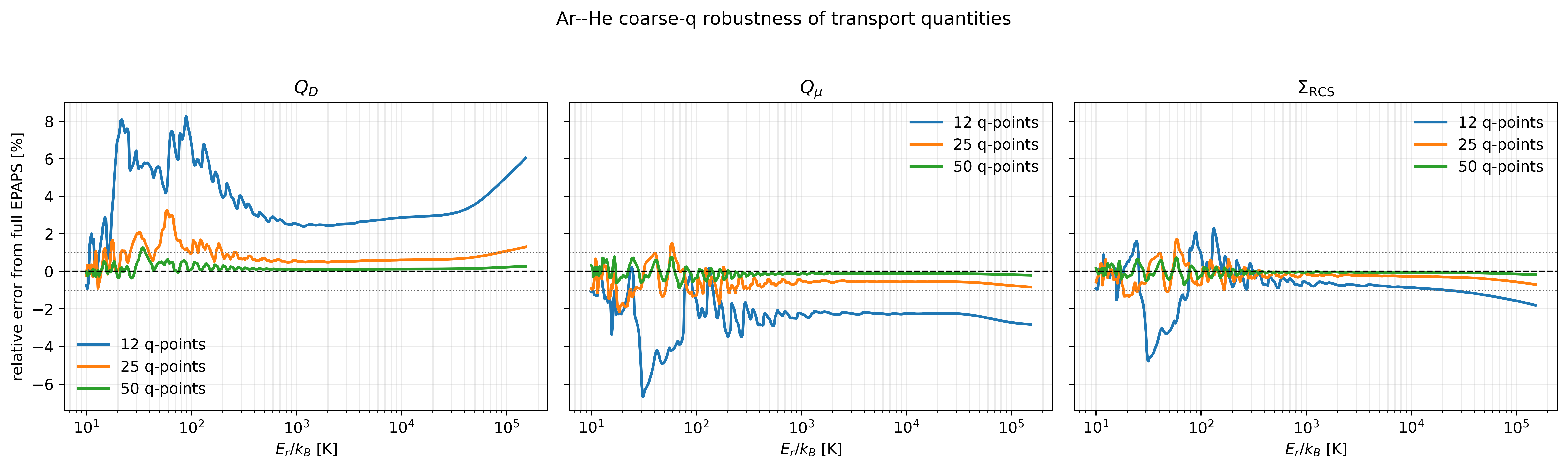}
\includegraphics[width=0.96\textwidth,trim=0 0 0 0.45in,clip]{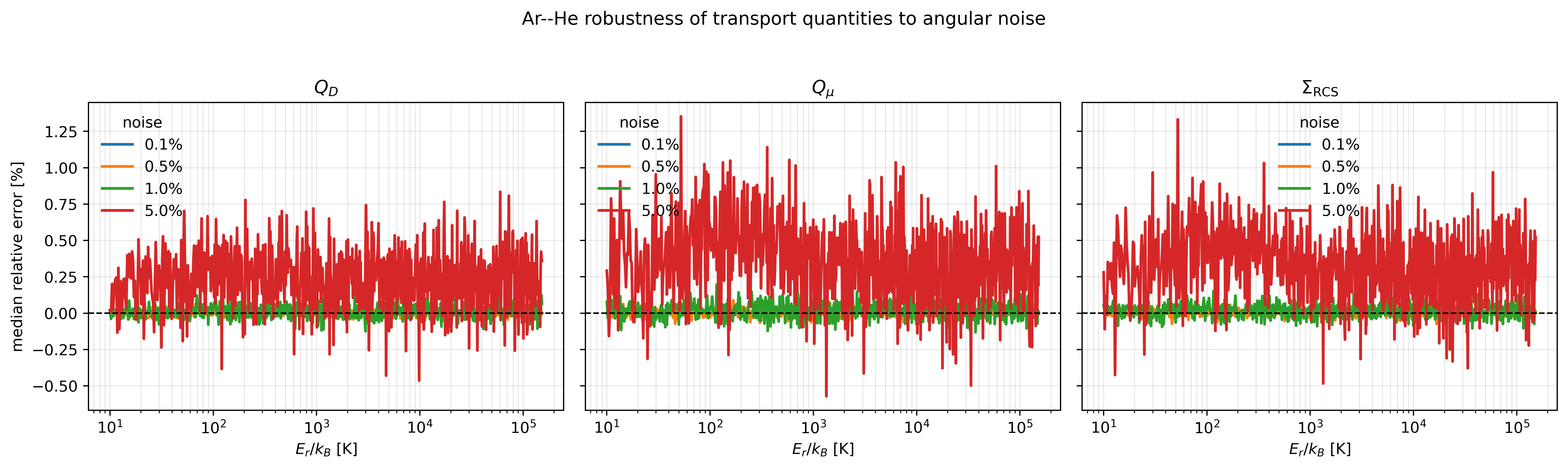}
\caption{Robustness tests inspired by coarse/noisy-input validation in flow reconstruction.  Top: transport errors induced by reconstructing the EPAPS map from coarser $q$ grids.  Bottom: median transport errors under synthetic angular noise applied to $\chi(q,E)$.}
\label{fig:coarse_noise}
\end{figure*}

A practical DSMC collision table cannot have infinite resolution.  Figure~\ref{fig:coarse_noise} therefore asks how much transport error is introduced if the EPAPS map is reconstructed from coarser $q$ grids.  The physical reason for the sensitivity is that the relevant integrals are area averages of nonlinear functions of $\chi$.  If a narrow range of impact area contains side-scattering or a rapid branch transition, a coarse quadrature can miss that area even if the plotted deflection curve appears qualitatively correct.  With only 12 $q$ points, $\QD$ errors can exceed $8\%$, confirming that under-resolving the impact-area coordinate is dangerous.  With 25 points, the errors drop substantially.  With 50 points, the maximum errors for $\Er/\kb\ge10~\mathrm{K}$ are $1.27\%$ for $\QD$, $0.88\%$ for $\Qmu$, and $0.73\%$ for $\RCS$.  This test explains why equal-area resolution is not a cosmetic detail: it controls the accuracy of transport projections.

The angular-noise test perturbs $\chi(q,E)$ by Gaussian noise proportional to the row-wise RMS of the angle.  Even for 5\% angular noise, the median transport errors remain about $0.83\%$ for $\QD$, $1.35\%$ for $\Qmu$, and $1.33\%$ for $\RCS$.  The reason is that uncorrelated angular perturbations partly cancel under the impact-area integral, whereas coherent displacement of a branch or systematic loss of high-mode content does not.  Thus, random noise and structured interpolation error are not equivalent.  This is not a claim that the physical scattering angle is noisy; rather, it quantifies how robust the transport projections are to perturbations that mimic fitting, interpolation, or finite-data errors.

\begin{table*}[t]
\centering
\caption{Summary of He--Ar neural-scattering validation metrics.  Errors are reported for $\Er/\kb\ge10~\mathrm{K}$ unless otherwise stated.  The cumulative-measure entry is reported as $100\max|\Delta\Sigma|$ for compact comparison with percent errors.}
\label{tab:summary}
\begin{tabular}{lll}
\toprule
Validation category & Metric & Deviation \\
\midrule
Pointwise angular map & median relative $L_2(\chi)$ & $0.34\%$ \\
Pointwise angular map & maximum relative $L_2(\chi)$ & $3.62\%$ \\
Spectral preservation & median high-$k_q$ ratio error & $0.008\%$ \\
Transport cross section & $\QD$ & $0.75\%$ \\
Transport cross section & $\Qmu$ & $1.37\%$ \\
Transport ratio & $\Qmu/\QD$ & $0.84\%$ \\
Ohr representative cross section & $\RCS$ & $1.21\%$ \\
Equivalent VSS cross section & $\SigVSS$ & $1.46\%$ \\
Angular cumulative measure & $\max |\Delta\Sigma(\mu)|$ & $1.43\%$ \\
Coarse impact-area grid & $\QD$ with 50 $q$ points & $1.27\%$ \\
Coarse impact-area grid & $\Qmu$ with 50 $q$ points & $0.88\%$ \\
Coarse impact-area grid & $\RCS$ with 50 $q$ points & $0.73\%$ \\
Angular-noise robustness & $\QD$ with 5\% angular noise & $0.83\%$ \\
Angular-noise robustness & $\Qmu$ with 5\% angular noise & $1.35\%$ \\
Angular-noise robustness & $\RCS$ with 5\% angular noise & $1.33\%$ \\
\bottomrule
\end{tabular}
\end{table*}

\subsection{DSMC validation: sinusoidal Ar--He relaxation}\label{sec:sinusoidal_test}

Before presenting the three solver-level tests, we specify the collision-step comparison used throughout them.  The DSMC simulations use the same cellwise no-time-counter/majorant-frequency pair-selection logic for the native EPAPS and neural runs.  For each species pair, the candidate-pair rate is based on a conservative majorant of $g\,\sigma_T(E_r)$ in the cell, and the same majorants, time step, simulator weights, number density, and sampling protocol are used in the paired EPAPS and NN calculations.  The Ar--Ar and He--He pair kernels are identical in the two runs.  For the He--Ar pair, the total scattering area $\sigma_T(E_r)$ and accepted-pair sampling of the equal-area coordinate $q$ are held fixed by the native EPAPS table; only the accepted-pair deflection law is changed from $\chi_{\EPAPS}(q,E_r)$ to $\chi_{\NN}(q,E_r)$.  Thus the reported solver-level differences isolate the learned angular scattering map rather than a change in the collision-rate closure.

The preceding sections validate the neural scattering map as a function and as a set of transport and measure projections.  A remaining question is whether these kernel-level agreements survive when the learned He--Ar scattering law is placed inside a DSMC mixture calculation.  To address this point without introducing the additional complications of walls, external forcing, or multidimensional particle noise, we use a periodic one-dimensional binary-mixture relaxation problem.  The test is deliberately minimal: it isolates cross-species diffusion and directly compares the native EPAPS He--Ar kernel with the neural He--Ar kernel while keeping the same Ar--Ar and He--He pair tables, initial particle realization, number density, temperature, domain size, and sampling protocol.

The initial helium mole fraction is prescribed as a small sinusoidal perturbation about an equimolar mixture,
\begin{equation}
  X_{\rm He}(x,0)=X_0+A_0\sin(kx),\qquad X_0=0.5,
  \qquad k=\frac{2\pi}{L},
  \label{eq:sinusoidal_initial}
\end{equation}
with periodic boundaries.  For small perturbations and nearly uniform temperature and pressure, the first composition mode should relax diffusively,
\begin{equation}
  A(t)\simeq A_0\exp(-k^2D_{12}t),
  \label{eq:sinusoidal_decay}
\end{equation}
where $D_{12}$ is the effective binary mutual-diffusion coefficient represented by the collision kernel.  In the DSMC data, the mode amplitude is evaluated directly from particle labels rather than from a noisy binned mole-fraction profile,
\begin{equation}
  A(t)=2\left\langle \left[I_{\rm He,p}(t)-X_0\right]\sin\left(\frac{2\pi x_p(t)}{L}\right)\right\rangle_p,
  \label{eq:particle_mode_amplitude}
\end{equation}
where $I_{\rm He,p}=1$ for a helium simulator and zero otherwise.  This particle-level estimator reduces binning noise and gives a robust scalar diagnostic of composition relaxation.  We compare the normalized histories using
\begin{equation}
  \varepsilon_A=\frac{\left\|A_{\NN}(t)/A_{\NN}(0)-A_{\EPAPS}(t)/A_{\EPAPS}(0)\right\|_2}
  {\left\|A_{\EPAPS}(t)/A_{\EPAPS}(0)\right\|_2},
  \label{eq:history_error}
\end{equation}
computed over the early-time signal-carrying interval.  This metric is less sensitive than a single exponential fit to the late-time regime where the mode has decayed into DSMC statistical noise.

\begin{figure*}[t]
\centering
\includegraphics[width=0.92\textwidth,trim=0 45 0 0,clip]{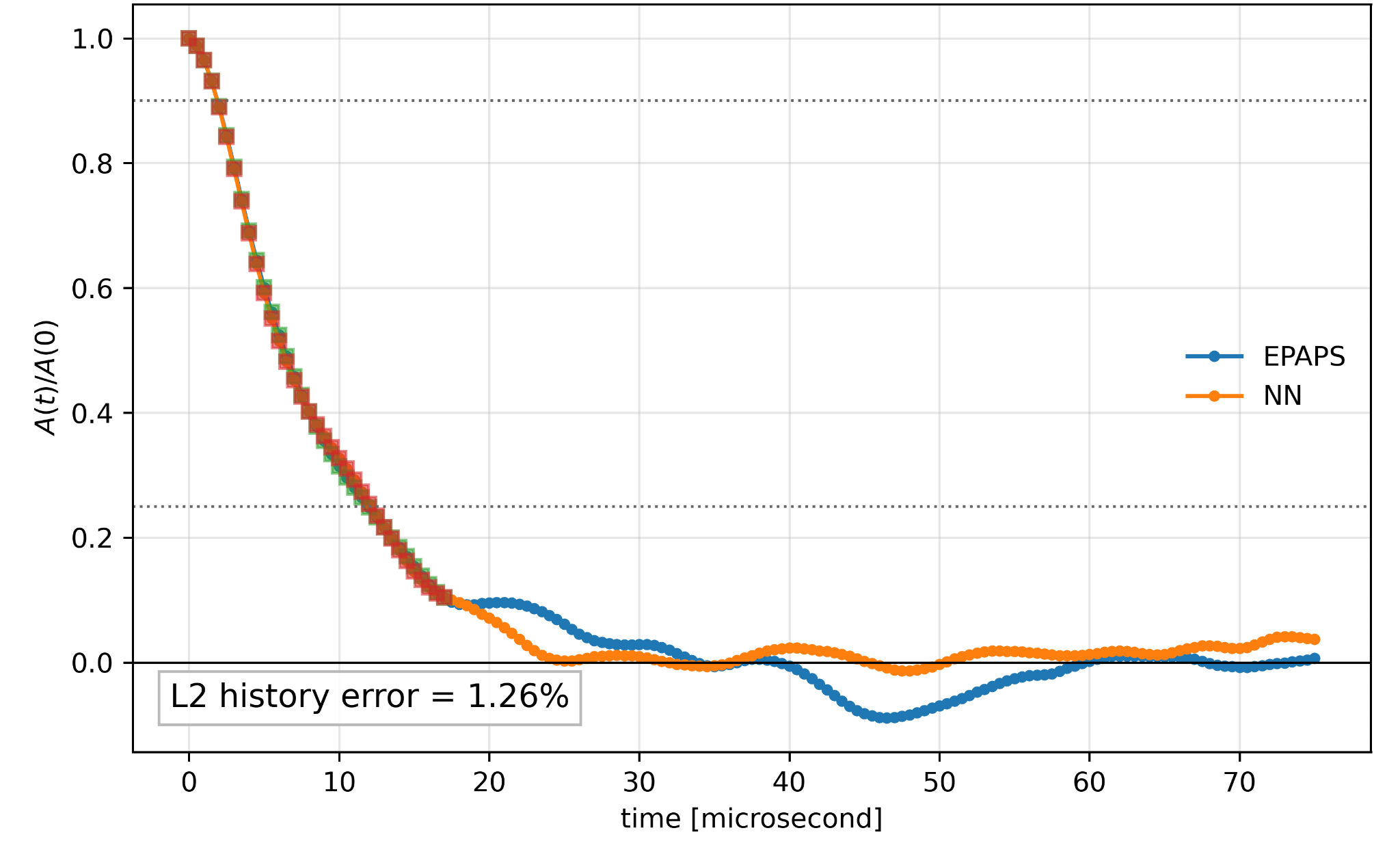}
\caption{Representative binary-mixture DSMC validation using periodic Ar--He sinusoidal mutual-diffusion relaxation.  The native EPAPS and neural He--Ar scattering kernels are embedded in otherwise identical DSMC calculations with the same initial particle realization.  This realization gives a relative history error of $1.26\%$; three independent realizations give $1.28\pm0.22\%$ and $D_{\NN}/D_{\EPAPS}=1.015\pm0.013$.}
\label{fig:sinusoidal_history}
\end{figure*}

Figure~\ref{fig:sinusoidal_history} shows one representative realization.  Physically, this test is the kinetic analogue of a single Fourier mode in Fickian mutual diffusion: the mode decays because He--Ar collisions erase the spatial correlation between species label and position, while Ar--Ar and He--He collisions thermalize each species without directly changing the composition contrast.  The decay rate is therefore a sensitive solver-level projection of the cross-species diffusion cross section rather than a visual comparison of mole-fraction profiles.  The EPAPS and neural kernels start from the same particle realization and nearly collapse over the physically meaningful decay interval.  For this realization the relative $L_2$ error of the normalized relaxation history is $1.26\%$, and the maximum absolute difference over that interval is $1.41\times10^{-2}$.  To separate surrogate error from Monte Carlo sampling variability, the calculation was repeated for three independent initial particle realizations.  The mean history error was $1.28\pm0.22\%$ (standard error), and a common-window exponential fit over $0.25<A(t)/A(0)<0.90$ gave $D_{\NN}/D_{\EPAPS}=1.015\pm0.013$.  Thus, the learned He--Ar kernel is not only accurate in isolated angular and transport diagnostics; it also reproduces the native EPAPS response in a genuine binary-mixture DSMC relaxation problem.  This test is intentionally different from earlier Lennard--Jones self- or tracer-diffusion demonstrations: here the dominant relaxation mechanism is the cross-species He--Ar collision kernel, and the three pair-specific tables are used together in a mixture calculation.

\subsection{DSMC validation: transverse shear-wave relaxation}\label{sec:shearwave_test}

The composition-mode test in Sec.~\ref{sec:sinusoidal_test} probes binary mass diffusion.  A complementary DSMC-level observable is momentum diffusion.  To test this second transport channel without introducing walls or accommodation-model uncertainty, we initialize the same equimolar periodic Ar--He mixture with a transverse shear wave,
\begin{equation}
  u_y(x,0)=U_0\sin(kx),\qquad k=\frac{2\pi}{L}.
  \label{eq:shear_initial}
\end{equation}
For a small-amplitude linear perturbation in a nearly uniform mixture, the first transverse-velocity mode relaxes as
\begin{equation}
  A_u(t)\simeq A_u(0)\exp(-\nu_{\rm mix}k^2t),
  \label{eq:shear_decay}
\end{equation}
where $\nu_{\rm mix}$ is the effective kinematic viscosity represented by the collision kernel.  The mode amplitude is measured directly from particles using a mass-weighted Fourier estimator,
\begin{equation}
\begin{aligned}
  A_u(t)&=\frac{2}{M}\sum_p m_p v_{y,p}(t)
  \sin\left(\frac{2\pi x_p(t)}{L}\right),\\
  M&=\sum_p m_p .
\end{aligned}
  \label{eq:shear_amplitude}
\end{equation}
This estimator targets the mixture momentum mode rather than the velocity of one species alone.  The test is therefore distinct from the composition relaxation: the composition mode is controlled primarily by species diffusion, whereas the shear mode is controlled by momentum diffusion and the viscosity-related angular projection.  Both calculations use the same initial particle realization, number density, temperature, periodic domain, Ar--Ar and He--He tables, and sampling protocol; only the He--Ar kernel is changed from native EPAPS to its neural representation.

\begin{figure*}[t]
\centering
\includegraphics[width=0.92\textwidth]{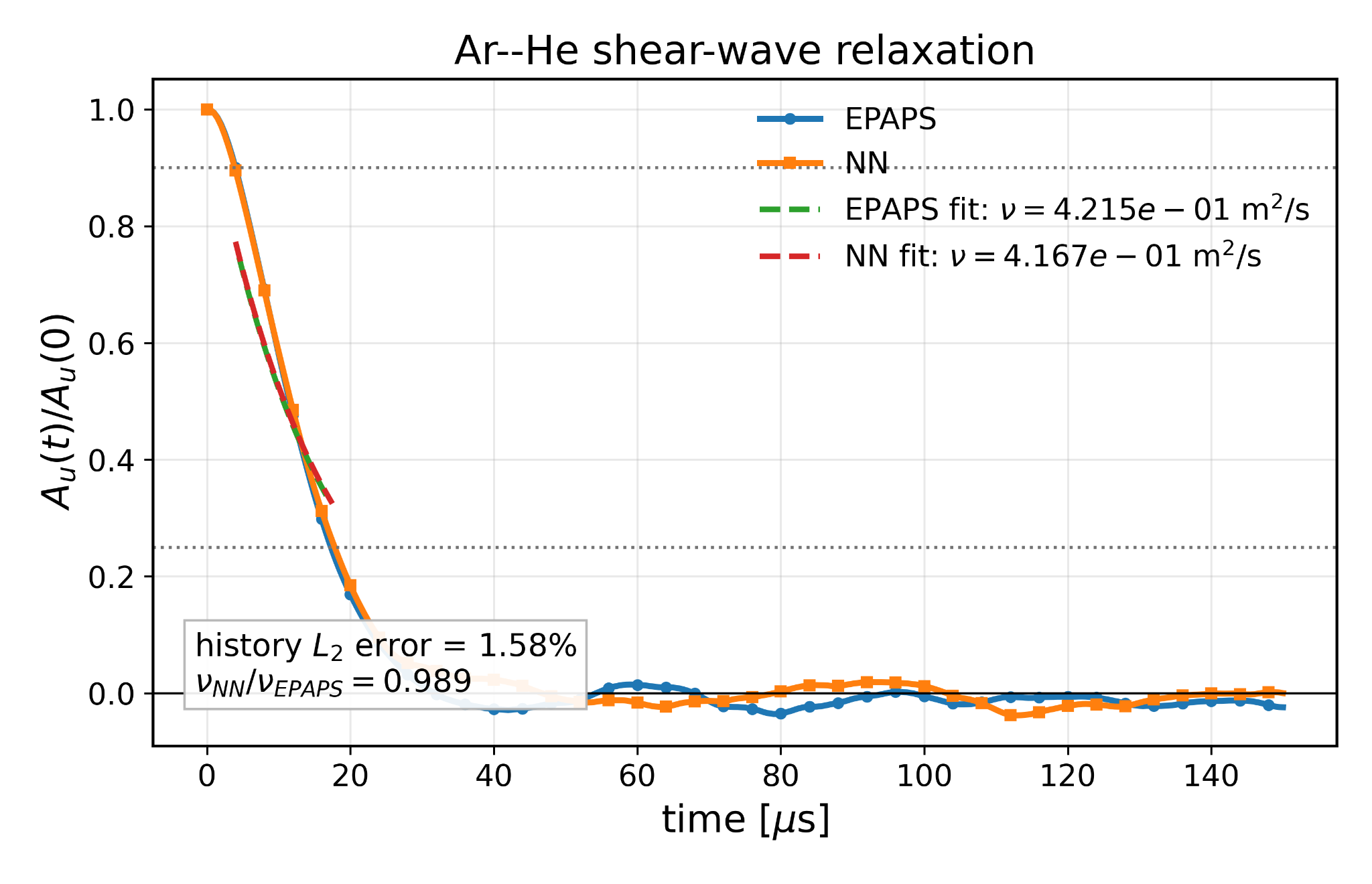}
\caption{Binary-mixture DSMC validation using periodic Ar--He transverse shear-wave relaxation.  The native EPAPS and neural He--Ar kernels are embedded in otherwise identical DSMC calculations.  The neural kernel reproduces the EPAPS momentum-mode relaxation with a relative history error of $1.58\%$.  A constrained exponential fit over $0.25<A_u(t)/A_u(0)<0.90$ gives $\nu_{\NN}/\nu_{\EPAPS}=0.989$.}
\label{fig:shearwave_history}
\end{figure*}

Figure~\ref{fig:shearwave_history} shows the normalized shear-mode relaxation.  This test is the momentum-transport counterpart of the composition mode.  The initial transverse velocity contains no wall forcing and no mean pressure gradient; it decays only because molecular collisions transfer transverse momentum between particles and relax the first Fourier mode of the shear stress.  It therefore emphasizes the viscosity-related angular projection $\Qmu$ and the interspecies momentum exchange controlled by the He--Ar kernel.  The EPAPS and neural histories are almost indistinguishable over the signal-carrying interval.  The relative history error is $1.58\%$ and the maximum absolute history difference is $1.76\times10^{-2}$.  A constrained exponential fit over $0.25<A_u(t)/A_u(0)<0.90$ gives
\begin{equation}
  \nu_{\EPAPS}=0.4215~\mathrm{m^2/s},\qquad
  \nu_{\NN}=0.4167~\mathrm{m^2/s},
  \label{eq:shear_nu_values}
\end{equation}
so that $\nu_{\NN}/\nu_{\EPAPS}=0.989$.  Thus, the neural He--Ar scattering kernel preserves not only the static viscosity cross-section projection $\Qmu$ but also the dynamic momentum-diffusion response of a DSMC mixture calculation.  Together with the composition-mode test, this result provides two independent solver-level checks: one for mass diffusion and one for momentum diffusion.

\subsection{DSMC validation: two-dimensional periodic shear-layer mixing}\label{sec:shearlayer_test}

The two single-mode DSMC tests above are intentionally clean: one isolates a scalar composition mode and the other isolates a transverse momentum mode.  A stricter solver-level test should nevertheless contain spatially evolving mixture structure.  We therefore add a two-dimensional periodic binary shear-layer problem.  The purpose is not to claim a new instability benchmark; rather, it is to test whether the neural He--Ar scattering kernel preserves the native EPAPS response in an unsteady, field-level mixture calculation that contains composition gradients, velocity gradients, scalar dissipation, and interspecies slip without any wall-accommodation model.  The periodic setting is important because it keeps the comparison focused on the collision kernel rather than on gas--surface scattering.

The initial mole-fraction and velocity fields are prescribed as
\begin{equation}
  X_{\rm He}(x,y,0)=0.5+0.4\tanh\left[\frac{y_i(x)-y}{\delta_X}\right],
  \qquad
  y_i(x)=\frac{L_y}{2}+a_i\sin\left(\frac{2\pi x}{L_x}\right),
  \label{eq:shearlayer_composition_initial}
\end{equation}
\begin{equation}
  u_x(y,0)=U_0\tanh\left(\frac{y-L_y/2}{\delta_u}\right),
  \qquad
  u_y(x,y,0)=\eps_u U_0\sin\left(\frac{2\pi x}{L_x}\right),
  \label{eq:shearlayer_velocity_initial}
\end{equation}
with periodic boundaries in both directions.  The production calculation uses $L_x=L_y=7.5\times10^{-4}~\mathrm{m}$, $T=300~\mathrm{K}$, $n=2\times10^{21}~\mathrm{m^{-3}}$, $U_0=550~\mathrm{m/s}$, $a_i=0.10L_y$, $\delta_X=0.045L_y$, $\delta_u=0.07L_y$, and $7\times10^5$ simulators on an $80\times80$ grid.  These parameters place the full box at a low global Knudsen number while the composition layer remains a finite-gradient kinetic region.  The native EPAPS and neural He--Ar calculations again use the same initial particle realization and identical Ar--Ar and He--He pair kernels.

The primary scalar metric is the normalized composition variance,
\begin{equation}
  M(t)=\left\langle \left(X_{\rm He}(x,y,t)-0.5\right)^2\right\rangle_{\Omega},
  \label{eq:mixing_index_2d}
\end{equation}
which decreases as the binary mixture homogenizes.  We also monitor the field-level mole-fraction error and the species-slip magnitude,
\begin{equation}
  S(t)=\left\langle \left|\mathbf{u}_{\rm He}(x,y,t)-\mathbf{u}_{\rm Ar}(x,y,t)\right|\right\rangle_{\Omega}.
  \label{eq:species_slip_metric}
\end{equation}
The slip field is especially relevant because it is a direct manifestation of interspecies momentum exchange; it is also statistically more demanding than $X_{\rm He}$ because it is obtained from the difference of two species-conditioned velocity averages.

\begin{figure*}[t]
\centering
\includegraphics[width=0.96\textwidth]{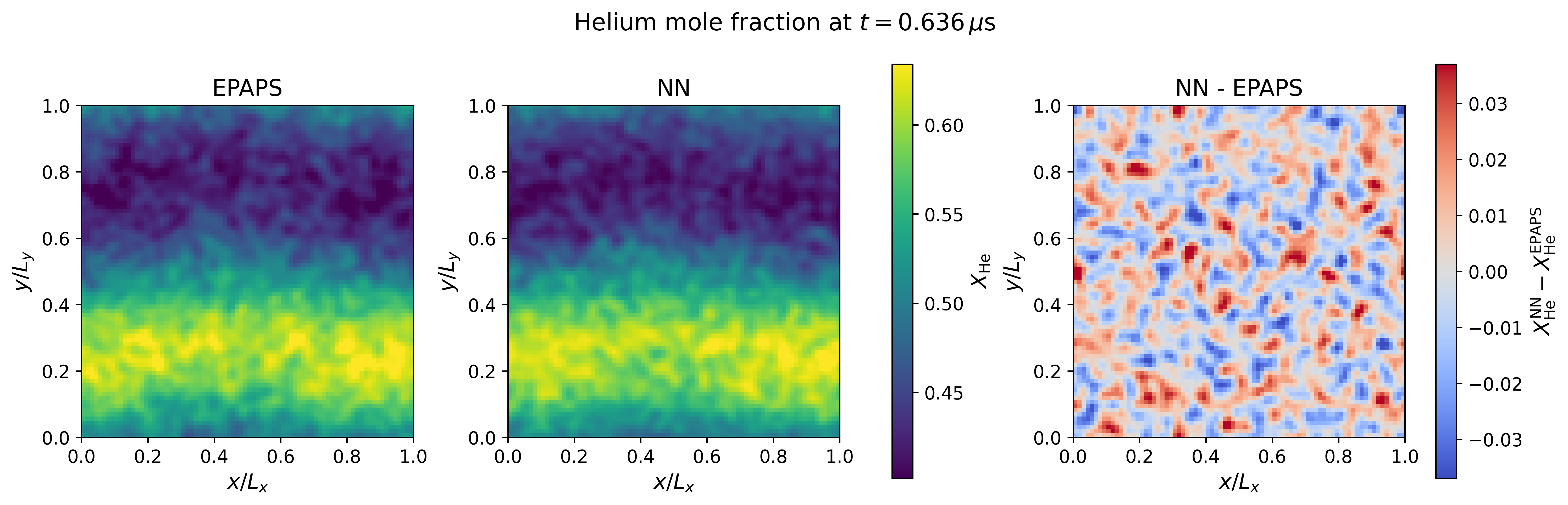}
\includegraphics[width=0.72\textwidth]{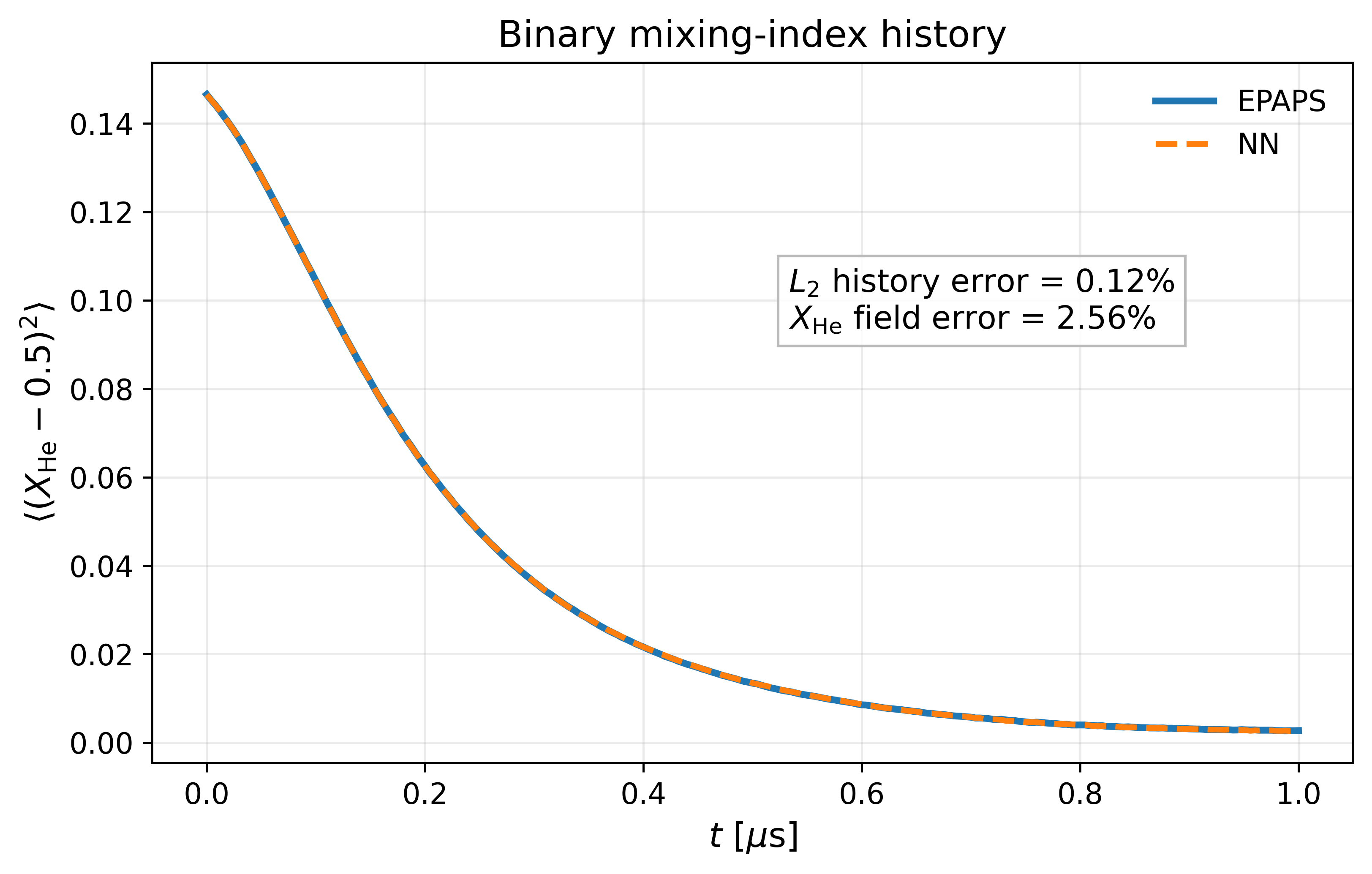}
\caption{Two-dimensional periodic Ar--He shear-layer validation.  Top: helium mole fraction at $t=0.636~\mu\mathrm{s}$ from the native EPAPS kernel, the neural He--Ar kernel, and their difference.  Bottom: binary mixing-index history.  The neural kernel reproduces the EPAPS mixing history with a relative $L_2$ error of $0.124\%$ and the selected-time $X_{\rm He}$ field with a relative $L_2$ error of $2.56\%$.}
\label{fig:shearlayer_main}
\end{figure*}

Figure~\ref{fig:shearlayer_main} shows the selected-time mole-fraction field and the complete mixing-index history.  The physics is different from the two modal tests.  The imposed shear convects the composition interface, increases interfacial length, and generates scalar gradients, while molecular diffusion and interspecies momentum exchange broaden the layer and reduce the variance $M(t)$.  The competition between advective stretching and collisional smoothing is what makes this a field-level kinetic test rather than a one-dimensional diffusion fit.  The neural and native EPAPS fields show the same broadened, sheared mixing layer.  The difference field is small and spatially structureless compared with the main composition band, indicating that the neural kernel does not introduce a systematic displacement or artificial sharpening of the interface.  Quantitatively, the normalized $M(t)$ history differs by only $0.124\%$ in relative $L_2$ norm, and the selected-time $X_{\rm He}$ field differs by $2.56\%$ in relative $L_2$ norm.  Thus the agreement seen in the one-dimensional composition and shear-wave tests persists in a two-dimensional evolving mixture field.

A Chapman--Enskog calculation provides a useful transport-theory scale for interpreting the field-level result.  From the EPAPS diffusion cross section $\QD(E)$, the first-order binary diffusion coefficient can be written in the present notation as
\begin{equation}
  D_{12}^{CE}(T)=\frac{3\pi}{8n\,\overline{Q}_D(T)}
  \left(\frac{2\kb T}{\pi m_{12}}\right)^{1/2},
  \qquad
  \overline{Q}_D(T)=\int_0^\infty \QD(x\kb T)x^2e^{-x}\,\dd x,
  \label{eq:CE_D12}
\end{equation}
where $m_{12}$ is the reduced mass.  At $T=300~\mathrm{K}$ and $n=2\times10^{21}~\mathrm{m^{-3}}$, this gives $D_{12}^{CE}=0.916~\mathrm{m^2/s}$.

\begin{figure*}[t]
\centering
\includegraphics[width=0.96\textwidth]{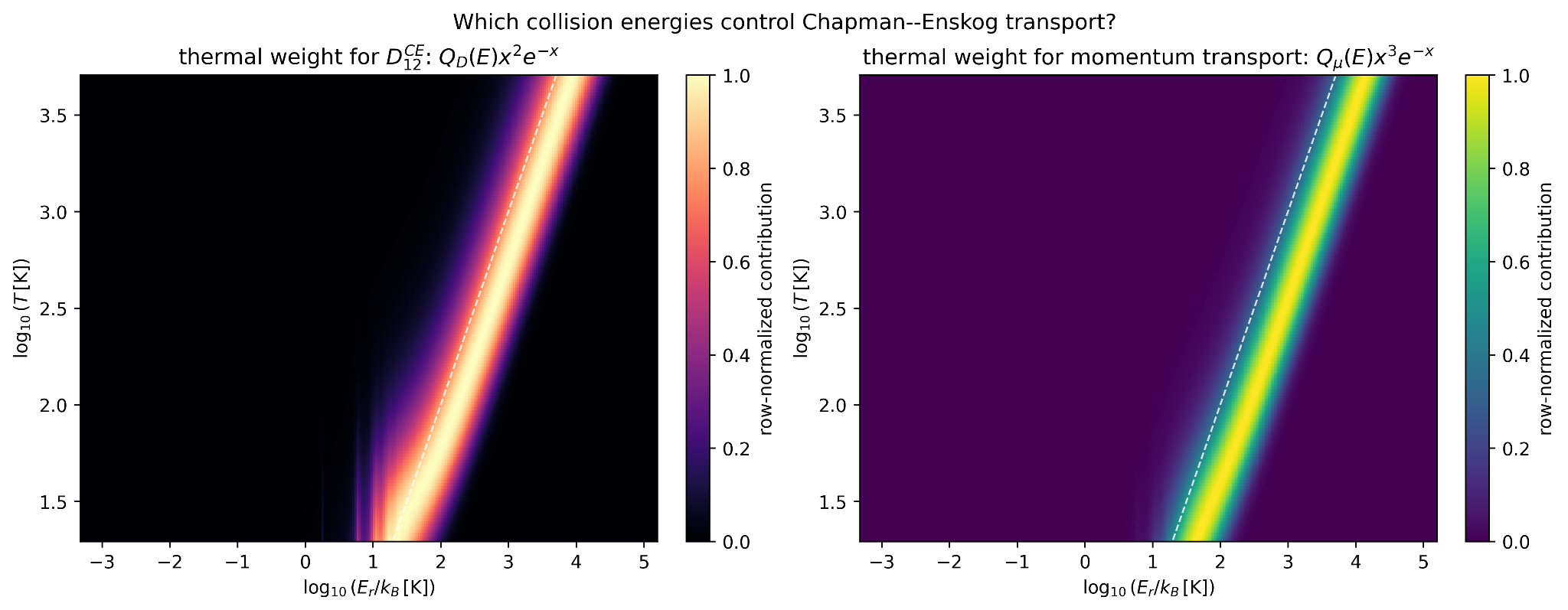}
\caption{Thermal contribution windows for Chapman--Enskog transport computed from the EPAPS He--Ar table.  The left panel shows the row-normalized contribution $Q_D(E)x^2e^{-x}$ controlling the binary diffusion integral, while the right panel shows the corresponding momentum-transport contribution $Q_\mu(E)x^3e^{-x}$.  The dashed diagonal marks $E_r/k_B=T$.  The finite-width diagonal bands show that a gas at temperature $T$ samples only a limited collision-energy window, with the momentum-transport window shifted to higher energy than the diffusion window.}
\label{fig:ce_thermal_contribution_maps}
\end{figure*}

The CE integral is not controlled uniformly by all tabulated collision energies.  Figure~\ref{fig:ce_thermal_contribution_maps} shows the row-normalized integrands $Q_D(E)x^2e^{-x}$ and $Q_\mu(E)x^3e^{-x}$, where $x=E_r/k_BT$.  These maps provide the physical bridge between the scattering table and the macroscopic transport coefficients.  For any temperature, the Boltzmann factor suppresses very high collision energies, while the powers of $x$ suppress the lowest energies; the result is a finite diagonal energy window that shifts upward with $T$.  The momentum-transport window is displaced to somewhat larger $E_r/k_B$ than the diffusion window because the viscosity-like projection carries an additional factor of $x$.  This explains why the low-energy operating envelope must be reported honestly but should not be judged by an unweighted maximum error over the whole table: a room-temperature DSMC calculation samples the $1$--$100~\mathrm{K}$ region only through the tail of the thermal contribution, whereas cryogenic simulations would weight that region much more strongly.  Thus the relevant accuracy is simultaneously energy-weighted by the collision distribution and angle-weighted by the transport projection.

Table~\ref{tab:temperature_weighted_errors} turns this observation into a quantitative operating-envelope metric.  Although the unweighted low-energy diagnostics in Appendix~\ref{app:additional_checks} contain localized maxima, the thermally weighted transport-projection errors remain below $0.12\%$ for $Q_D$, $Q_\mu$, and $\Sigma_{\rm RCS}$ over the representative range $20$--$3000~\mathrm{K}$.  The cumulative angular-measure and high-mode spectral metrics are larger because they probe the full angular push-forward and branch-sensitive high-$q$ structure, but they also decrease as the thermal collision window moves away from the most structured low-energy ridges.  The peak contribution to the diffusion integral lies above $T$ because the factor $x^2e^{-x}$ emphasizes relative energies of order a few $k_BT$, and the momentum-transport contribution is shifted still higher by the additional power of $x$.  This table therefore provides a more application-relevant accuracy statement than an unweighted maximum over the full EPAPS table.

\begin{table*}[t]
\centering
\small
\caption{Temperature-weighted He--Ar surrogate errors.  The entries are obtained by weighting the EPAPS--NN discrepancies by the thermal collision-energy contribution to the Chapman--Enskog transport integrals, rather than by taking an unweighted maximum over the whole scattering table.  The peak-energy column gives the maximum of the diffusion-weight contribution $Q_D(E)x^2e^{-x}$, with $x=E_r/k_B T$.  The cumulative-measure and spectral entries are reported as temperature-weighted mean absolute deviations.}
\label{tab:temperature_weighted_errors}
\begin{tabular}{ccccccc}
\toprule
$T$ [K] & peak $E_r/k_B$ [K] & $\epsilon_T(Q_D)$ [\%] & $\epsilon_T(Q_\mu)$ [\%] & $\epsilon_T(\Sigma_{\rm RCS})$ [\%] & $100\langle|\Delta\Sigma|\rangle_T$ [\%] & high-$k_q$ err. [\%] \\
\midrule
20 & 23.5 & 0.095 & 0.054 & 0.065 & 1.62 & 1.99 \\
77 & 135.3 & 0.061 & 0.078 & 0.080 & 1.19 & 0.82 \\
300 & 546.4 & 0.046 & 0.090 & 0.077 & 1.08 & 0.28 \\
1000 & 1807.7 & 0.053 & 0.112 & 0.084 & 1.04 & 0.16 \\
3000 & 5311.5 & 0.022 & 0.070 & 0.072 & 1.01 & 0.16 \\
\bottomrule
\end{tabular}
\end{table*}

\begin{figure*}[t]
\centering
\includegraphics[width=0.96\textwidth]{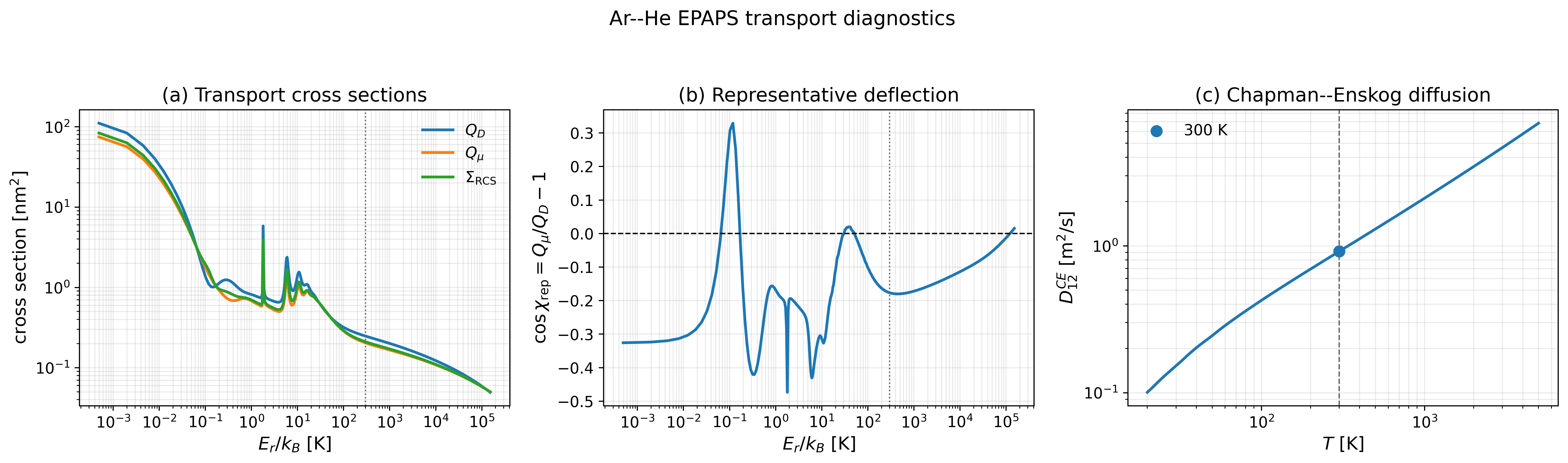}
\includegraphics[width=0.94\textwidth]{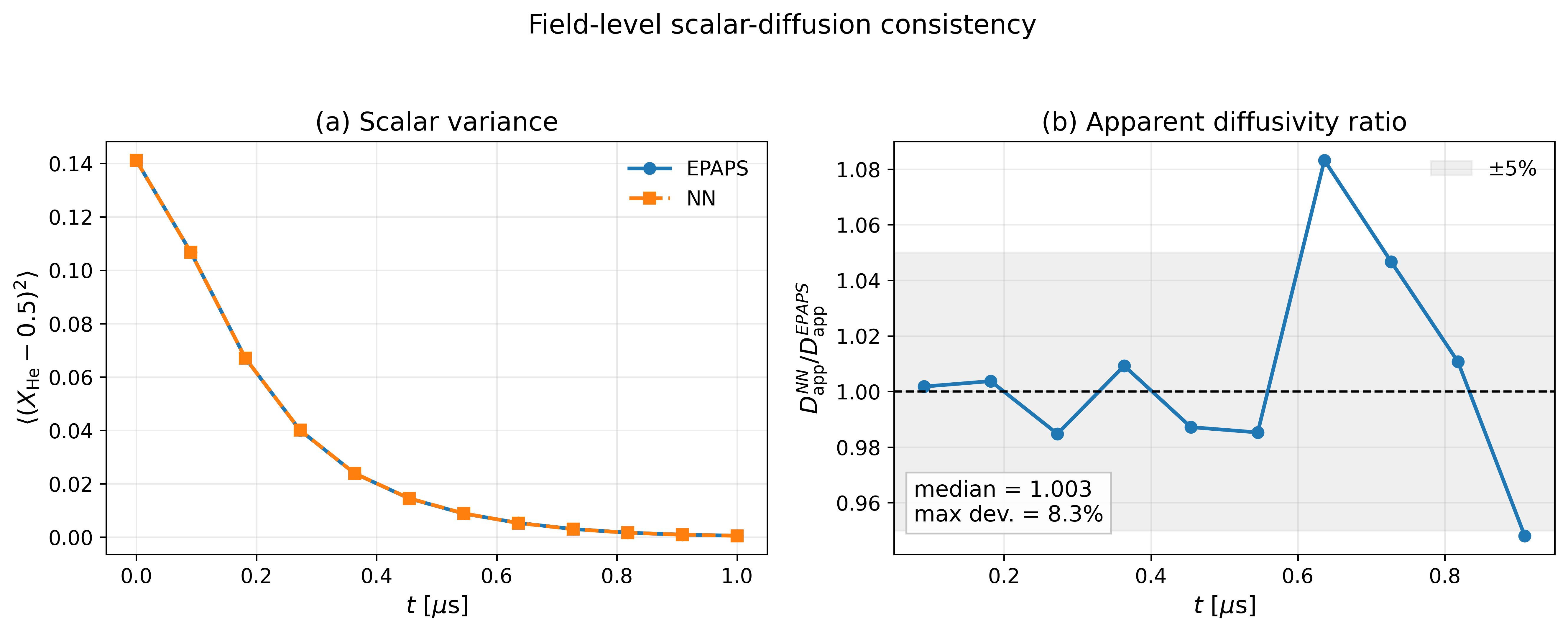}
\caption{Transport-theory interpretation of the two-dimensional shear-layer test.  Top: EPAPS-derived transport cross sections, representative-deflection diagnostic, and Chapman--Enskog binary diffusion scale for Ar--He.  Bottom: ratio of apparent scalar diffusivities extracted from the DSMC variance budget.  The absolute $D_{\rm app}$ value is a field-level apparent scale in a sheared finite-gradient problem; the NN-to-EPAPS ratio is the relevant surrogate-preservation metric.}
\label{fig:shearlayer_CE}
\end{figure*}

Figure~\ref{fig:shearlayer_CE} also reports the Ohr-style transport cross sections and representative-deflection diagnostic obtained from the same EPAPS table.  To compare the sheared DSMC field with this transport scale, we define an apparent scalar diffusivity from the scalar-variance budget,
\begin{equation}
  D_{\rm app}(t)=-\frac{1}{2}
  \frac{\dd M/\dd t}{\left\langle |\nabla X_{\rm He}|^2\right\rangle_{\Omega}}.
  \label{eq:Dapp}
\end{equation}
Because the shear-layer problem contains coherent advection, finite gradients, nonlinear mixing, and DSMC sampling noise, $D_{\rm app}$ should not be overinterpreted as the homogeneous Chapman--Enskog coefficient itself.  The robust validation quantity is instead the ratio $D_{\rm app}^{\NN}/D_{\rm app}^{\EPAPS}$, which asks whether the neural kernel preserves the apparent scalar-diffusion response of the native EPAPS calculation.  Over the resolved interval, the median ratio is $1.003$, with a mean absolute deviation of about $2.5\%$ and a maximum deviation of about $8.3\%$.  This comparison ties the two-dimensional field test back to the EPAPS transport integrals while avoiding an inappropriate equality between a sheared finite-gradient DSMC field and a homogeneous CE experiment.

The same production run also reproduces the mean composition and streamwise-velocity profiles, the scalar-dissipation proxy, and the mean species-slip history; these are reported in Appendix~\ref{app:additional_checks}.  The selected-time species-slip field has a larger relative error ($6.37\%$) than the mole-fraction field, as expected for a difference of species-conditioned velocities, but it remains close at the level of the dominant spatial structure.  This slip is not a numerical artifact: in a binary gas with a large mass ratio, He and Ar can carry different conditional velocities inside the mixing layer before cross-species collisions relax their relative drift.  It is therefore a direct field-level measure of interspecies momentum exchange.  Vorticity is more sensitive still because it differentiates a noisy DSMC velocity field; it is therefore treated as a qualitative supplementary diagnostic rather than as a primary validation metric.

\begin{table*}[t]
\centering
\small
\caption{Solver-level DSMC mixture-mode validation.  The composition-mode entries report the mean and standard error over three independent particle realizations.  The shear-wave entries are from the paired EPAPS/NN realization shown in Fig.~\ref{fig:shearwave_history}.  The shear-layer entries are from the production two-dimensional paired EPAPS/NN run in Fig.~\ref{fig:shearlayer_main}.  An EPAPS-only particle/cell refinement check is reported in Appendix~\ref{app:additional_checks}.}
\label{tab:dsmc_modes}
\begin{tabular}{lccc}
\toprule
DSMC test & Primary transport channel & History error & Fitted rate ratio \\
\midrule
Sinusoidal composition mode & binary mass diffusion & $1.28\pm0.22\%$ & $D_{\NN}/D_{\EPAPS}=1.015\pm0.013$ \\
Transverse shear wave & momentum diffusion / viscosity & $1.58\%$ & $\nu_{\NN}/\nu_{\EPAPS}=0.989$ \\
2D periodic shear layer & field-level mixing / slip & $0.124\%$ & $D_{\rm app}^{\NN}/D_{\rm app}^{\EPAPS}=1.003$ \\
\bottomrule
\end{tabular}
\end{table*}

Table~\ref{tab:dsmc_modes} collects the three solver-level mixture tests in the order used in the text: sinusoidal mass-diffusion relaxation, transverse momentum-diffusion relaxation, and two-dimensional field-level shear-layer mixing.  The table is therefore the solver-level counterpart of the pair-level summary in Table~\ref{tab:summary}: it reports whether preservation of the He--Ar angular measure and transport projections remains visible after the surrogate is embedded in DSMC mixture dynamics.

\section{Discussion}\label{sec:discussion}

The results in Table~\ref{tab:summary} help answer an important objection: if the Sharipov--Benites EPAPS table already exists, what is gained by passing it through a neural network?  The gain is not new ab initio physics.  The reference physics remains the ab initio potential and the scattering data.  The gain also should not be overclaimed as unconditional storage reduction or automatic wall-clock acceleration relative to every high-quality interpolation strategy.  Dense non-neural interpolation, especially PCHIP on a carefully chosen energy--impact grid, can be a strong competitor for a single fixed pair potential.  The more defensible advantage of the neural representation is that it provides a smooth, continuously evaluable, differentiable, and DSMC-ready scattering map whose physical admissibility can be checked through transport, measure, representative, spectral, and solver-level diagnostics.  A neural fit that only minimizes angle error would indeed be little more than interpolation.  A transport-, measure-, representative-, and spectrum-preserving surrogate is better interpreted as a validated collision-kernel representation.  The interpolation baselines in Appendix~\ref{app:additional_checks} are therefore used to quantify the accuracy--storage--smoothness tradeoff rather than to claim that the neural table dominates every possible direct interpolation baseline.  The interpolation and benchmark diagnostics in Tables~\ref{tab:baseline_interp} and~\ref{tab:runtime_memory} make this point explicit: dense PCHIP is a strong accuracy--storage baseline for this single fixed table, the exported neural table has lookup throughput comparable to the dense native table, compact linear tables are smaller and slightly faster in this vectorized benchmark, and online PCHIP evaluation trades storage efficiency for much lower direct evaluation throughput unless it is pretabulated or otherwise optimized.

The hierarchy also identifies which aspects of the map are easy and which are difficult.  Median pointwise angular error and median high-mode spectral error are very small.  Transport errors are below about $1.5\%$ over the DSMC-relevant range.  The difficult region is $1$--$100~\mathrm{K}$, where the attractive part of the He--Ar potential gives long collision times, large effective scattering areas, and high-curvature deflection branches.  In that interval, the cross sections are not smooth monotone functions of energy because $\QD$ and $\Qmu$ are area integrals over a deflection function that can change rapidly with impact parameter.  This is why the spectral, cumulative-measure, and representative-cross-section diagnostics are physically informative rather than merely statistical.  This low-energy interval should not be hidden as an outlier; it is the operational boundary of the present surrogate.  At cryogenic temperatures, a large fraction of the collision-energy distribution can sample precisely these ridges, whereas at room temperature and above the Chapman--Enskog contribution maps show that the dominant collision-energy window shifts toward higher energies.  The relevant error is therefore the error weighted by the thermal collision-energy distribution and by the transport projection, not only the unweighted maximum over the full table.  The coarse-grid test shows that using too few impact-area points can produce several-percent errors, whereas 50 points are already sufficient to keep the main transport projections near the one-percent level.  This suggests that a neural surrogate should not be judged against a single dense table alone; it should also be tested against deliberate coarsening, perturbation, and temperature-weighted operating windows.

The Ohr-style metrics provide a useful bridge between full angular scattering and reduced collision models.  Because $\RCS$, $\cos\RDA$, and $\alpha_{\rm VSS}$ are nonlinear functions of $\QD$ and $\Qmu$, they test correlated errors in the transport projections.  Their preservation means that the neural map would give nearly the same representative single-deflection reduction as the reference EPAPS map.  This is particularly useful for DSMC implementations where collision-rate and angular-sampling strategies may be separated.

The spectral validation connects the present scattering-kernel problem to recent findings in data-driven turbulence sensing.  In both cases, pointwise MSE alone is insufficient.  A model can look accurate in physical space but lose important high-frequency content; conversely, a model can preserve integrated quantities while still distorting fine-scale structure.  The Fourier spectrum of $\chi(q,E)$ is not a transport coefficient by itself, but it diagnoses whether the learned map preserves high-curvature and branch-sensitive features.  The loss-ablation study reinforces this point: a small transport-aware penalty improves held-out angular generalization, but stronger or simultaneous measure penalties do not automatically improve the model.  Thus, physics-aware loss design is useful, but it must be treated as a tuned kinetic-learning problem rather than as a monotone addition of constraints.

The three DSMC mixture tests complete this validation ladder by embedding the learned cross-species kernel in transport dynamics.  The first two tests are intentionally simple and modal: no accommodation model, imposed wall flux, or multidimensional wake noise is needed to interpret the mass- and momentum-relaxation rates.  The scalar composition mode is controlled primarily by binary mutual diffusion, whereas the transverse shear mode is controlled by momentum diffusion.  The two-dimensional shear-layer test then adds spatially evolving mixture structure while keeping periodic boundaries, so that the comparison remains focused on the collision kernel rather than on gas--surface scattering.  The three-realization composition error of $1.28\pm0.22\%$ in Table~\ref{tab:dsmc_modes}, the $1.58\%$ shear-history error in Fig.~\ref{fig:shearwave_history}, and the $0.124\%$ two-dimensional mixing-history error in Fig.~\ref{fig:shearlayer_main} show that the angular, transport, measure, and spectral agreements survive in mass diffusion, momentum diffusion, and field-level binary mixing.  In all three tests the Ar--Ar and He--He kernels are held fixed while the He--Ar map is changed from EPAPS to its neural surrogate.  This provides a stronger answer to the interpolation objection: the surrogate is not only a compact fit to a table, but a collision kernel that reproduces the mass-, momentum-, and field-level mixing behavior of the native ab initio data.

Additional checks, summarized in Appendix~\ref{app:additional_checks}, address practical concerns that are not central enough to appear as main figures.  First, non-neural interpolation baselines show that coarse direct interpolation can produce larger transport errors than the neural representation, although dense PCHIP interpolation can approach comparable accuracy.  Second, an EPAPS-only shear-wave refinement gives a total coarse-to-fine spread of only $1.8\%$ in the fitted momentum-diffusion rate, indicating that the NN--EPAPS shear discrepancy is not dominated by particle/cell resolution.  Third, a low-energy operating-envelope analysis identifies the $1$--$100~\mathrm{K}$ interval as the most sensitive part of the map while showing that median transport-projection errors in that band remain below $0.3\%$.

\section{Conclusions}\label{sec:conclusion}

We developed a multiscale validation framework for neural ab initio scattering kernels intended for DSMC gas-mixture simulations.  The main conclusion is that pointwise deflection-angle error is necessary but not sufficient.  A DSMC-ready neural collision kernel must preserve the angular push-forward measure and the transport projections that determine diffusion, viscosity, representative collision rates, and angular redistribution; it should also reproduce independent mixture-transport responses when inserted into a DSMC calculation, including a field-level two-dimensional mixing response.

The framework was applied to a refined Ar--Ar J\"ager map and to a He--Ar EPAPS neural surrogate.  A five-model training ablation showed that the best held-out angular validation loss was obtained with a modest transport-aware penalty, $\lambda_Q=0.05$, giving a $14.4\%$ reduction relative to angle-only training; larger transport weighting and the tested cumulative-measure penalties degraded this validation metric.  For He--Ar and $\Er/\kb\ge10~\mathrm{K}$, the selected surrogate preserved $\QD$, $\Qmu$, $\Qmu/\QD$, $\RCS$, and $\SigVSS$ within $0.75\%$, $1.37\%$, $0.84\%$, $1.21\%$, and $1.46\%$, respectively.  The cumulative angular measure agreed within about $1.4\%$ in the normalized metric reported here.  Spectral analysis showed that the median relative $L_2$ error of $\chi(q)$ is $0.34\%$ and that the median high-mode spectral-energy ratio is essentially unity.  Coarse-grid and angular-noise tests confirmed that transport projections remain controlled under finite-resolution and perturbed scattering data.

Three final DSMC tests embedded the learned He--Ar map in periodic binary-mixture dynamics.  In the composition-mode test, the neural and native EPAPS kernels were run from the same initial particle realization with identical Ar--Ar and He--He pair kernels.  Across three independent realizations, the normalized first-mode composition history from the neural kernel matched the EPAPS history with a relative $L_2$ error of $1.28\pm0.22\%$, and a common-window exponential fit gave $D_{\NN}/D_{\EPAPS}=1.015\pm0.013$.  In the transverse shear-wave test, the neural kernel reproduced the EPAPS momentum-mode history with a relative $L_2$ error of $1.58\%$, and a constrained exponential fit gave $\nu_{\NN}/\nu_{\EPAPS}=0.989$.  In the two-dimensional periodic shear-layer test, the neural kernel reproduced the EPAPS mixing-index history with a $0.124\%$ error, the selected-time helium mole-fraction field with a $2.56\%$ relative $L_2$ error, and the species-slip field with a $6.37\%$ error.  A CE calculation from the EPAPS diffusion cross section gave $D_{12}^{CE}=0.916~\mathrm{m^2/s}$ for the production density and temperature; the apparent scalar-diffusion ratio extracted from the sheared field had median $D_{\rm app}^{\NN}/D_{\rm app}^{\EPAPS}=1.003$.  These tests show that kernel-level preservation of angular measures and transport projections transfers to mixture responses for mass diffusion, momentum diffusion, and two-dimensional field-level mixing.

These results support a validation philosophy for learned kinetic collision kernels: they should be accepted not because they reduce regression error, but because they preserve the measure, transport, spectral, and relaxation structure of the reference scattering physics over the collision-energy window sampled by the intended application.  The present evidence is strongest for elastic noble-gas pair scattering and for the He--Ar mixture tests reported here.  Extension to other potentials, polyatomic gases, reactive or inelastic channels, and multicomponent mixtures should retain the same validation ladder but should not be assumed from the present pair alone.  In future DSMC implementations, pair-specific learned scattering maps should therefore be reported together with their collision-rate closure, temperature-weighted operating envelope, interpolation/runtime baseline, and solver-level uncertainty.

\appendix
\section{Additional robustness and numerical-sensitivity checks}\label{app:additional_checks}

This appendix collects compact numerical checks that support the main validation hierarchy but are secondary to the main physical narrative.  They address practical concerns: whether the neural representation is merely a weak interpolation baseline, how storage and evaluator throughput compare with direct interpolation strategies, where the remaining surrogate error lies when weighted by transport physics, whether the DSMC shear-wave result is sensitive to the particle/cell resolution, how the surrogate behaves in the low-energy interval where the scattering map is most structured, and how the two-dimensional shear-layer fields behave in species-slip and profile diagnostics.

\begin{table*}[h]
\centering
\small
\caption{Non-neural interpolation baselines for the He--Ar EPAPS map.  Errors are maximum absolute errors for $\Er/\kb\ge10~\mathrm{K}$.  The dense PCHIP baseline approaches comparable accuracy, but coarse interpolation can produce several-percent errors in the same transport projections.}
\label{tab:baseline_interp}
\begin{tabular}{lcccc}
\toprule
Representation & Storage [MB] & $\QD$ err. [\%] & $\Qmu$ err. [\%] & $\RCS$ err. [\%] \\
\midrule
Neural exported table & 0.967 & 0.75 & 1.37 & 1.21 \\
Linear, $75E\times25q$ & 0.014 & 3.02 & 5.52 & 3.70 \\
Linear, $150E\times50q$ & 0.057 & 1.64 & 2.29 & 1.85 \\
PCHIP, $300E\times50q$ & 0.114 & 1.04 & 0.85 & 0.77 \\
\bottomrule
\end{tabular}
\end{table*}

\begin{table*}[h]
\centering
\small
\caption{Storage and evaluation-throughput benchmark for different DSMC-ready He--Ar scattering representations.  Throughput was measured for vectorized table/evaluator calls in the Python benchmark script and is reported only as an implementation-level comparison, not as a hardware-independent limit.  The benchmark distinguishes compact storage from online evaluation cost: dense and exported linear tables have similar throughput, compact linear tables are slightly faster in this vectorized test, while direct online PCHIP evaluation is much slower unless it is pretabulated or optimized.}
\label{tab:runtime_memory}
\begin{tabular}{lccc}
\toprule
Representation/evaluator & Array storage [MB] & File size [MB] & Throughput [M eval/s] \\
\midrule
native dense linear lookup & 0.728 & 1.153 & 5.09 \\
neural exported table lookup & 0.728 & 1.014 & 5.20 \\
linear, $150E\times50q$ & 0.0616 & -- & 6.34 \\
linear, $300E\times50q$ & 0.123 & -- & 6.11 \\
PCHIP, $300E\times50q$ & 0.123$^\dagger$ & -- & 0.006 \\
\bottomrule
\end{tabular}
\vspace{0.2em}
\begin{flushleft}
\footnotesize $^\dagger$The PCHIP row uses the same stored knot table as the $300E\times50q$ compact representation; additional interpolation coefficients are implementation dependent and were not archived as a separate file in this benchmark.  Here, \emph{neural exported table lookup} denotes the neural surrogate sampled back onto a DSMC-ready equal-area table, not live neural-network inference.
\end{flushleft}
\end{table*}

\begin{table*}[h]
\centering
\small
\caption{EPAPS-only shear-wave refinement check.  The same constrained exponential fit over $0.25<A_u/A_u(0)<0.90$ was used for all resolutions.  The total coarse-to-fine spread is $1.8\%$ of the mean fitted value.}
\label{tab:shear_refinement}
\begin{tabular}{lcccc}
\toprule
Resolution & Particles & Cells & $\nu_{\EPAPS}$ [$\mathrm{m^2/s}$] & $\nu/\nu_{\rm base}$ \\
\midrule
Coarse & $8.0\times10^4$ & 80 & 0.41396 & 0.982 \\
Baseline & $1.2\times10^5$ & 120 & 0.42146 & 1.000 \\
Fine & $1.6\times10^5$ & 160 & 0.41856 & 0.993 \\
\bottomrule
\end{tabular}
\end{table*}

Table~\ref{tab:shear_refinement} provides the numerical-resolution context for the shear-wave comparison in Fig.~\ref{fig:shearwave_history}.  The fitted EPAPS kinematic-viscosity scale changes by less than about two percent from the coarse to the fine particle/cell setting, and the baseline value used for the NN--EPAPS comparison lies between the two refinement levels.  This check does not replace a full multi-seed uncertainty analysis, but it shows that the reported shear-wave discrepancy is not an obvious artifact of the nominal grid or particle count.

\begin{table*}[h]
\centering
\small
\caption{Low-energy operating envelope of the He--Ar neural surrogate over $1\le \Er/\kb\le100~\mathrm{K}$.  The cumulative-measure entry is reported as $100\max|\Delta\Sigma(\mu)|$ for compact comparison with percent errors.}
\label{tab:low_energy_envelope}
\begin{tabular}{lccc}
\toprule
Metric & Maximum absolute deviation & Median absolute deviation & Mean absolute deviation \\
\midrule
$\QD$ & 1.77\% & 0.21\% & 0.27\% \\
$\Qmu$ & 3.42\% & 0.21\% & 0.31\% \\
$\RCS$ & 1.21\% & 0.18\% & 0.24\% \\
$\max|\Delta\Sigma(\mu)|$ & 8.00\% & 1.00\% & 1.60\% \\
High-$k_q$ spectral ratio error & 12.28\% & 0.86\% & 1.56\% \\
\bottomrule
\end{tabular}
\end{table*}

Tables~\ref{tab:baseline_interp} and~\ref{tab:runtime_memory} should be interpreted as accuracy--storage--runtime diagnostics rather than separate claims of universal optimality.  The neural representation is not asserted to dominate every possible interpolation strategy in storage, raw interpolation accuracy, or wall-clock speed.  In fact, the dense PCHIP baseline is a strong competitor for this single fixed He--Ar table and can be smaller than the exported neural table.  The runtime benchmark clarifies the practical tradeoff: compact linear tables are inexpensive and fast when the table is already exported, whereas online PCHIP evaluation is much slower in the unoptimized vectorized Python implementation.  The practical value of the neural representation in this work is therefore the combination of smooth differentiability, continuous evaluation before export, compatibility with transport-aware training, and successful passage through transport, measure, spectrum, and DSMC relaxation checks.  The low-energy analysis also clarifies the main limitation: isolated energies in the $1$--$100~\mathrm{K}$ interval show larger angular-measure sensitivity than the median behavior, so this interval should be monitored explicitly when the surrogate is used in low-temperature DSMC applications.

\begin{figure*}[t]
\centering
\includegraphics[width=0.92\textwidth]{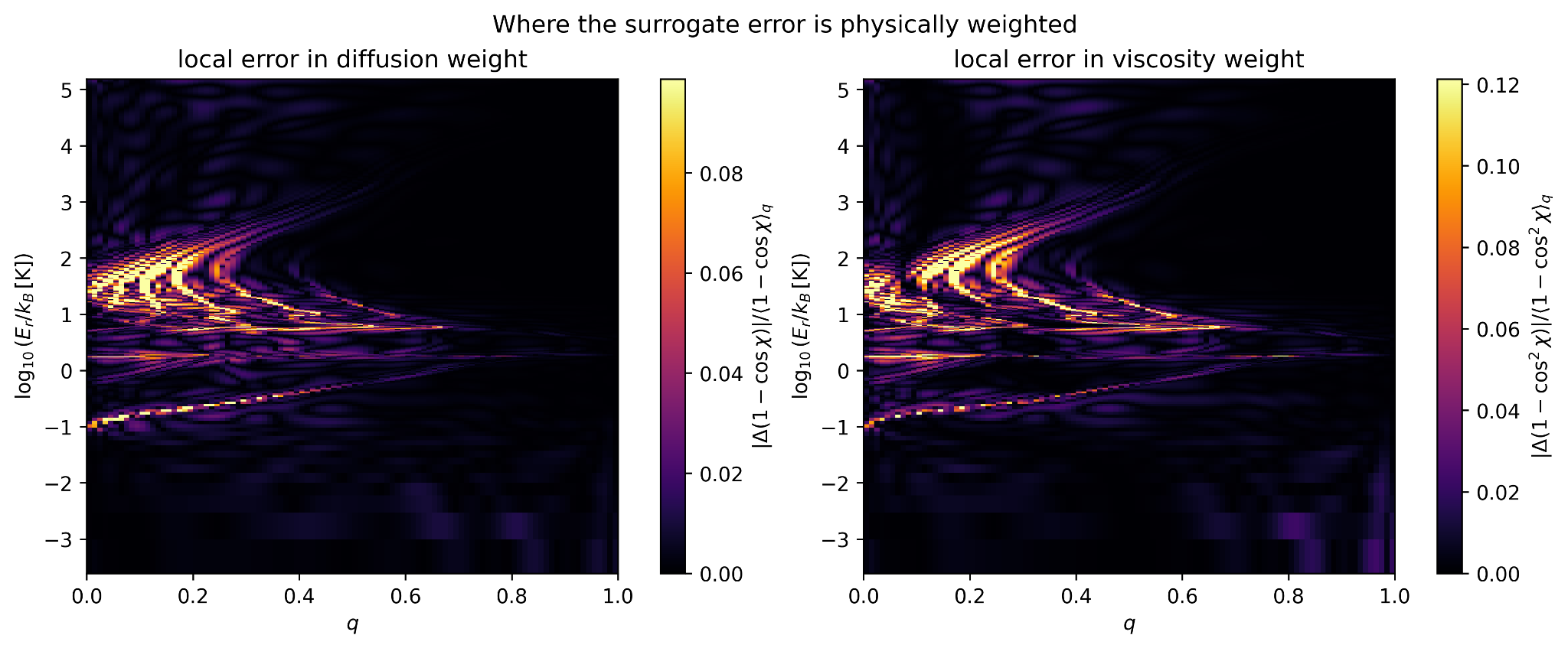}
\caption{Transport-weighted local error surfaces for the He--Ar neural surrogate.  The left panel weights the local discrepancy by the diffusion kernel $1-\cos\chi$, and the right panel uses the viscosity kernel $1-\cos^2\chi$.  The dominant residuals are localized along branch-sensitive low- and intermediate-energy ridges rather than distributed uniformly over the table.  This supports the operating-envelope interpretation of the low-energy statistics.}
\label{fig:appendix_transport_weighted_error}
\end{figure*}

Figure~\ref{fig:appendix_transport_weighted_error} provides the local counterpart of the global interpolation and low-energy statistics.  It shows the magnitude of the NN--EPAPS error after weighting by the diffusion and viscosity kernels.  The strongest residuals are not spread over the entire scattering table; they are confined to the same low- and intermediate-energy ridges where the deflection angle changes rapidly with impact area.  This is useful for interpreting Table~\ref{tab:low_energy_envelope}.  The larger maximum errors in the low-energy operating envelope come from localized branch-sensitive regions, whereas the median errors remain small because most of the impact-area/energy domain is smooth and forward focused.  The weighted-error maps therefore turn the low-energy limitation into an operational diagnostic: if a DSMC application thermally samples these ridges, the table should be refined or retrained there; if not, their influence on transport is strongly reduced by the thermal and angular weights.

\begin{figure*}[t]
\centering
\includegraphics[width=0.96\textwidth]{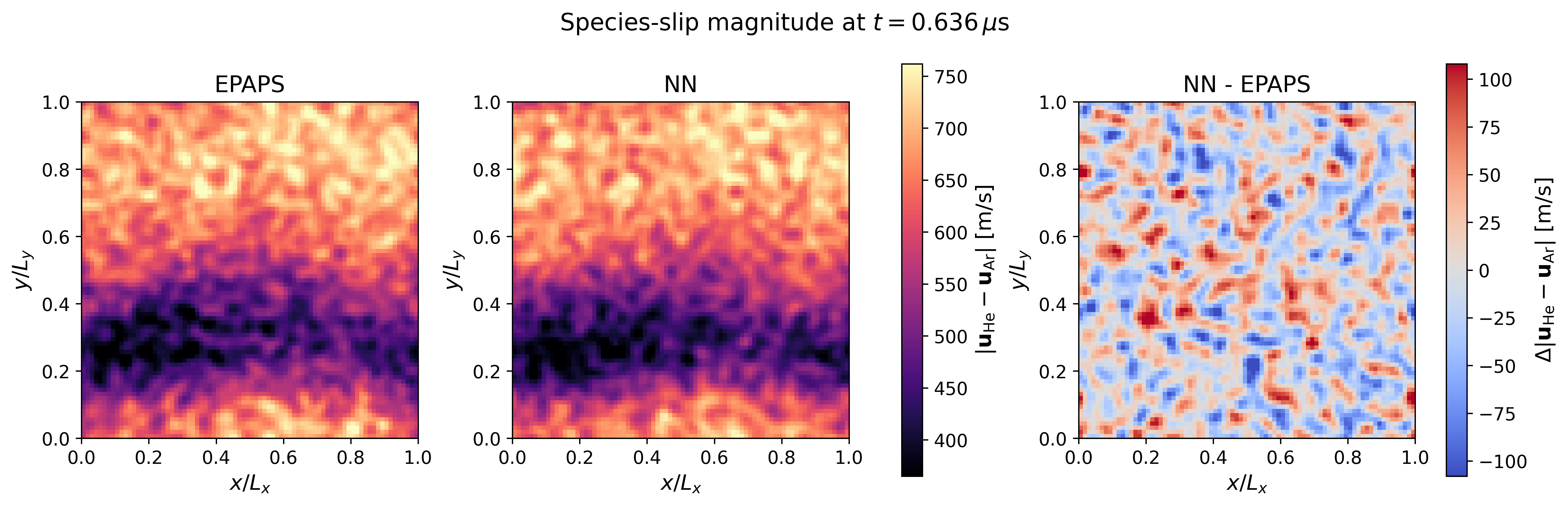}
\includegraphics[width=0.72\textwidth]{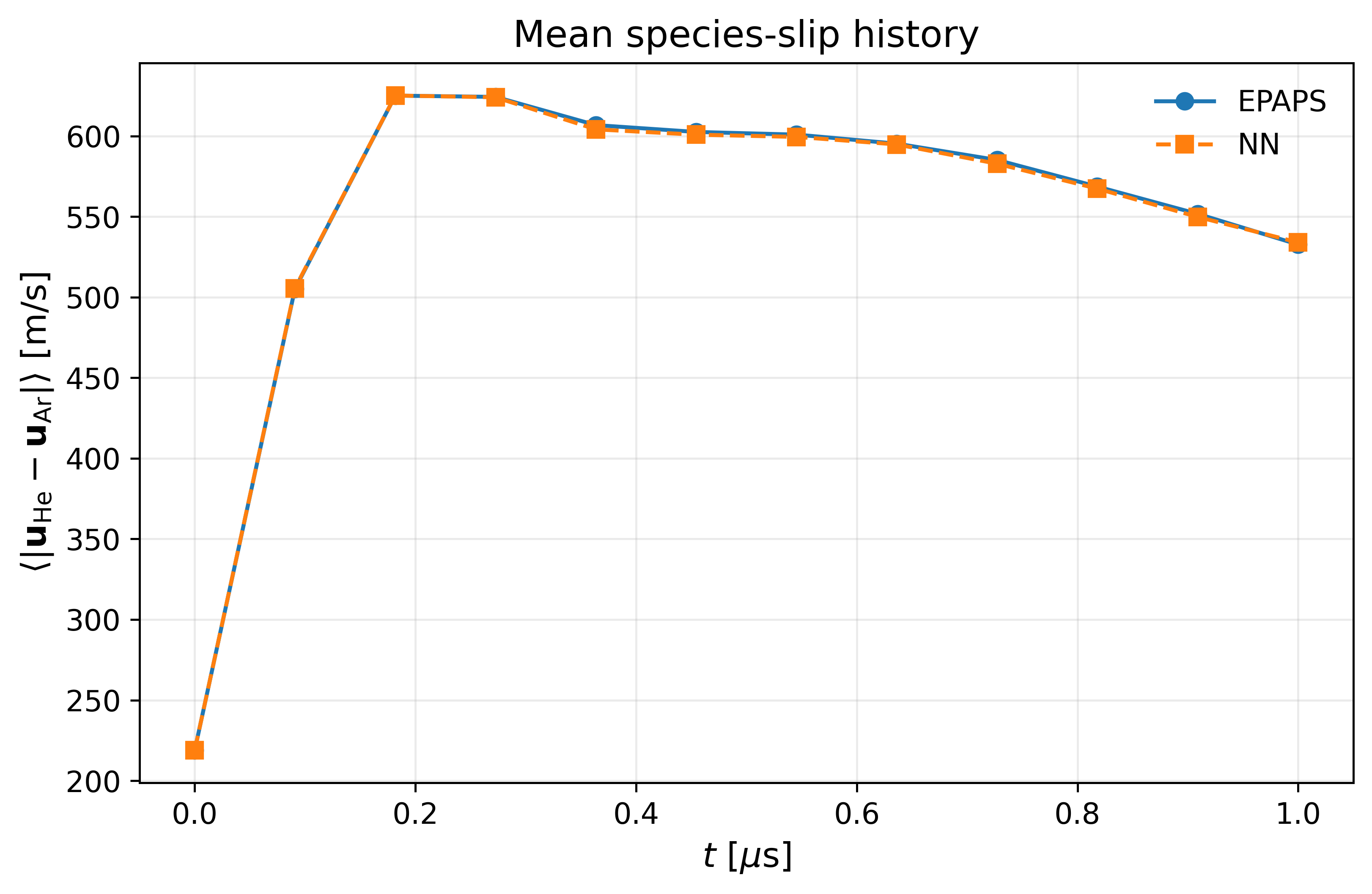}
\caption{Supplementary species-slip diagnostics for the two-dimensional periodic shear-layer validation.  Top: selected-time slip magnitude from EPAPS, NN, and their difference.  Bottom: domain-averaged species-slip history.  The slip metric is more sampling-sensitive than $X_{\rm He}$ because it is obtained from species-conditioned velocity differences, but the dominant spatial structure and temporal magnitude are preserved.}
\label{fig:appendix_slip}
\end{figure*}

\begin{figure*}[t]
\centering
\includegraphics[width=0.78\textwidth]{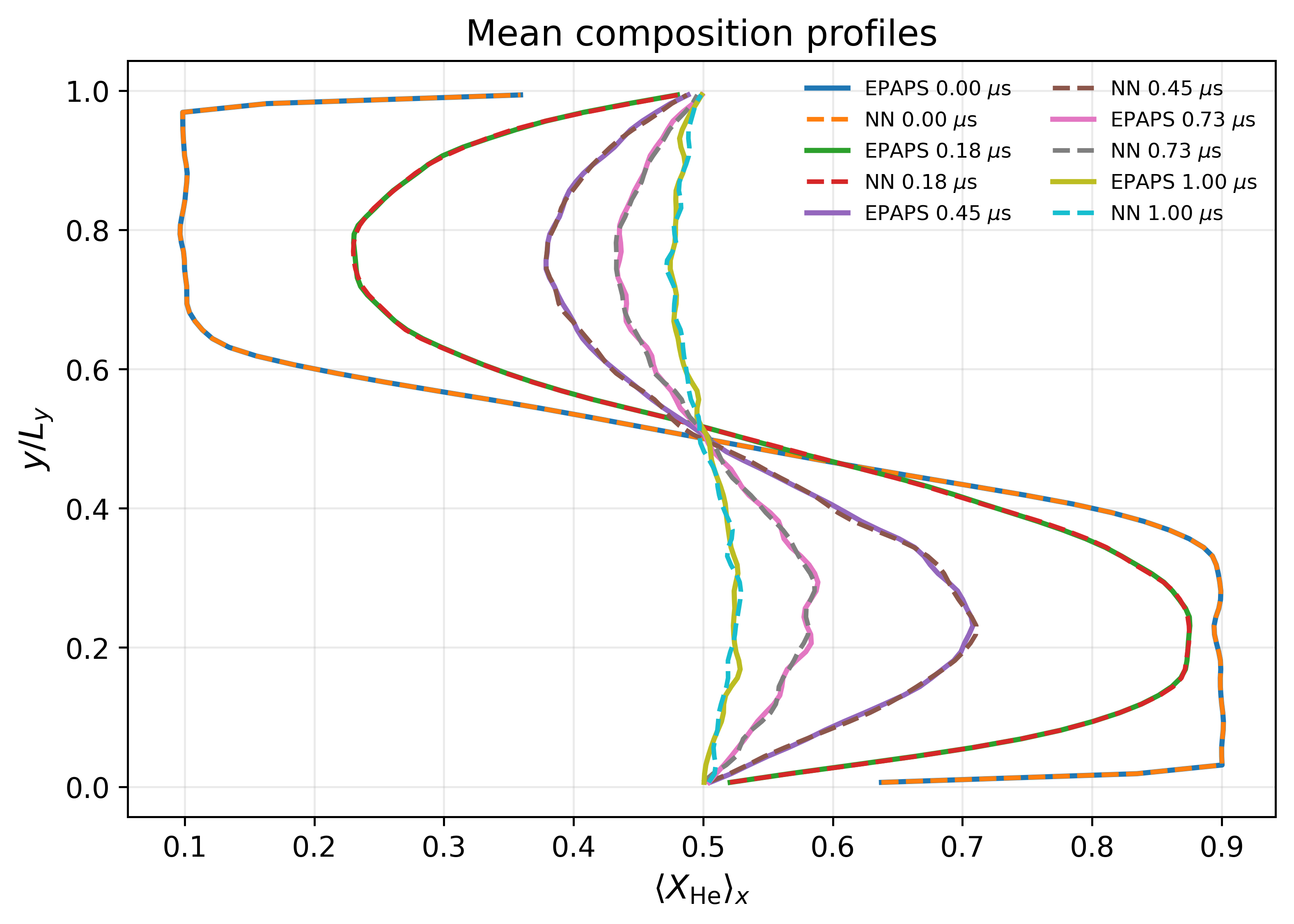}
\includegraphics[width=0.78\textwidth]{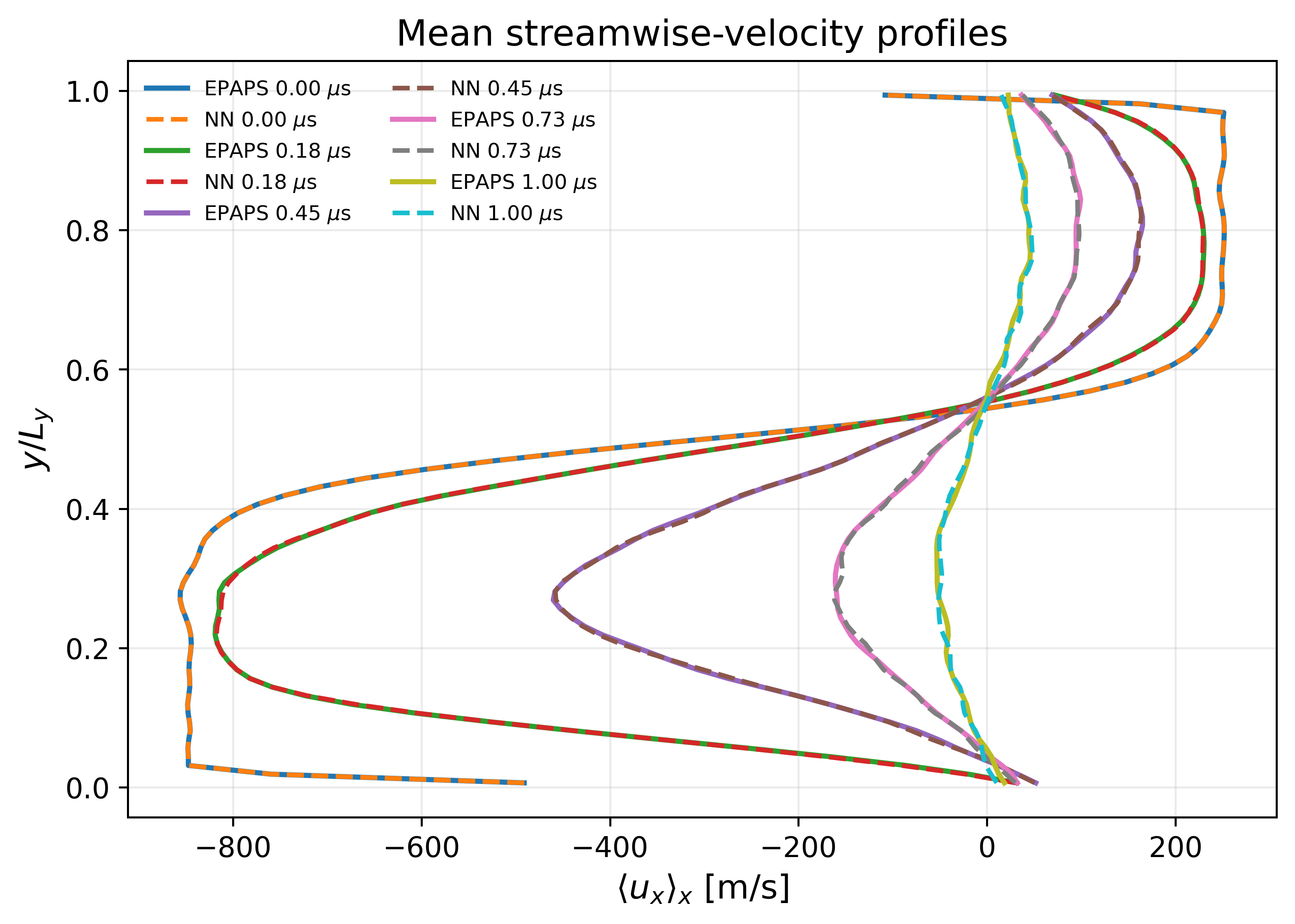}
\caption{Mean one-dimensional profiles extracted from the two-dimensional shear-layer production run.  Top: streamwise-averaged helium mole-fraction profiles.  Bottom: streamwise-averaged streamwise-velocity profiles.  Solid lines denote native EPAPS and dashed lines denote the neural He--Ar kernel at the same snapshot times.}
\label{fig:appendix_profiles}
\end{figure*}

Figure~\ref{fig:appendix_slip} complements the mole-fraction validation in Fig.~\ref{fig:shearlayer_main} by examining the species-slip magnitude, which is a more noise-sensitive but physically direct measure of local interspecies momentum exchange.  Figure~\ref{fig:appendix_profiles} then reduces the two-dimensional production run to streamwise-averaged composition and velocity profiles; these profiles confirm that the NN and EPAPS calculations preserve the same large-scale layer broadening and mean momentum relaxation, not only the domain-integrated mixing index.

\clearpage
\section*{Data availability}
The processed Ar--Ar, He--He, and He--Ar scattering tables used in this manuscript, the exported He--Ar neural table, validation summaries, DSMC composition-, shear-wave-, and two-dimensional shear-layer relaxation data, and the plotting scripts used to generate the figures will be provided to the editors and reviewers upon request through a private review-access repository.  The public version of the repository will be archived with a persistent DOI upon publication and will contain the processed tables, trained weights or exported neural tables, run metadata, random seeds, and scripts needed to regenerate the figures and tables reported here.  The He--Ar reference scattering data are derived from the EPAPS/ab initio dataset of Sharipov and Benites; any redistribution of the original third-party data will follow the permissions associated with that source, while the derived validation summaries and neural tables generated in this work will be archived with the manuscript.
\clearpage
\bibliographystyle{apsrev4-2}
\bibliography{references}

@article{Bird1963,
  author = {Bird, G. A.},
  title = {Approach to Translational Equilibrium in a Rigid Sphere Gas},
  journal = {Physics of Fluids},
  volume = {6},
  pages = {1518--1519},
  year = {1963},
  doi = {10.1063/1.1706763}
}

@article{Bird1970,
  author = {Bird, G. A.},
  title = {Direct Simulation and the Boltzmann Equation},
  journal = {Physics of Fluids},
  volume = {13},
  pages = {2676--2681},
  year = {1970},
  doi = {10.1063/1.1692849}
}

@article{Nanbu1980,
  author = {Nanbu, K.},
  title = {Direct Simulation Scheme Derived from the Boltzmann Equation. I. Monocomponent Gases},
  journal = {Journal of the Physical Society of Japan},
  volume = {49},
  pages = {2042--2049},
  year = {1980},
  doi = {10.1143/JPSJ.49.2042}
}

@article{KouraMatsumoto1991,
  author = {Koura, K. and Matsumoto, H.},
  title = {Variable Soft Sphere Molecular Model for Air Species},
  journal = {Physics of Fluids A},
  volume = {3},
  pages = {2459--2465},
  year = {1991},
  doi = {10.1063/1.858184}
}

@article{Aziz1995,
  author = {Aziz, R. A. and Janzen, A. R. and Moldover, M. R.},
  title = {Ab Initio Calculations for Helium: A Standard for Transport Property Measurements},
  journal = {Physical Review Letters},
  volume = {74},
  pages = {1586--1589},
  year = {1995},
  doi = {10.1103/PhysRevLett.74.1586}
}

@article{Bertoncini1970,
  author = {Bertoncini, P. J. and Wahl, A. C.},
  title = {Ab Initio Calculation of the Helium--Helium $^1\Sigma_g^+$ Potential Energy Curve},
  journal = {Physical Review Letters},
  volume = {25},
  pages = {991--993},
  year = {1970},
  doi = {10.1103/PhysRevLett.25.991}
}

@article{Vogel2010,
  author = {Vogel, Eckhard and J{"a}ger, Benjamin and Bich, Eckard and Hellmann, Robert},
  title = {Ab Initio Pair Potential Energy Curve for the Argon Atom Pair and Thermophysical Properties for the Dilute Argon Gas},
  journal = {Molecular Physics},
  volume = {108},
  pages = {3335--3352},
  year = {2010},
  doi = {10.1080/00268976.2010.507557}
}

@article{Jager2011,
  author = {J{"a}ger, Benjamin and Hellmann, Robert and Bich, Eckard and Vogel, Eckhard},
  title = {Ab Initio Virial Equation of State for Argon Using a New Nonadditive Three-body Potential},
  journal = {Journal of Chemical Physics},
  volume = {135},
  pages = {084308},
  year = {2011},
  doi = {10.1063/1.3626520}
}

@article{SharipovBenites2015,
  author = {Sharipov, Felix and Benites, Victor J.},
  title = {Transport Coefficients of Helium--Argon Mixture Based on Ab Initio Potential},
  journal = {Journal of Chemical Physics},
  volume = {143},
  pages = {154104},
  year = {2015},
  doi = {10.1063/1.4933327}
}

@article{SharipovBenites2019,
  author = {Sharipov, Felix and Benites, Victor J.},
  title = {Transport Coefficients of Argon and Its Mixtures with Helium and Neon at Low Density Based on Ab Initio Potentials},
  journal = {Fluid Phase Equilibria},
  volume = {498},
  pages = {23--32},
  year = {2019},
  doi = {10.1016/j.fluid.2019.06.010}
}

@article{Sharipov2020,
  author = {Sharipov, Felix and Benites, Victor J.},
  title = {Transport Coefficients of Multi-component Mixtures of Noble Gases Based on Ab Initio Potentials: Viscosity and Thermal Conductivity},
  journal = {Physics of Fluids},
  volume = {32},
  pages = {077104},
  year = {2020},
  doi = {10.1063/5.0016261}
}

@article{Sharipov2024,
  author = {Sharipov, Felix},
  title = {Ab Initio Modelling of Transport Phenomena in Multi-component Mixtures of Rarefied Gases},
  journal = {International Journal of Heat and Mass Transfer},
  volume = {220},
  pages = {124906},
  year = {2024},
  doi = {10.1016/j.ijheatmasstransfer.2023.124906}
}

@article{Ohr2023,
  author = {Ohr, Young Gie},
  title = {Representative Deflection Angle for the Single-deflection Method of Direct Simulation Monte Carlo},
  journal = {Physical Review E},
  volume = {108},
  pages = {035301},
  year = {2023},
  doi = {10.1103/PhysRevE.108.035301}
}

@article{Balasubramanian2026,
  author = {Balasubramanian, A. G. and Cremades, A. and Vinuesa, R. and Tammisola, O.},
  title = {Sharper Predictions: The Role of Loss Functions for Enhanced Turbulent-flow Sensing},
  journal = {Physical Review Fluids},
  volume = {11},
  pages = {044907},
  year = {2026},
  doi = {10.1103/26js-tpg4}
}

@book{RoohiAkhlaghiStefanov2025DSMCBook,
  author    = {Ehsan Roohi and Hassan Akhlaghi and Stefan Stefanov},
  title     = {Advances in Direct Simulation Monte Carlo: From Micro-Scale to Rarefied Flow Phenomena},
  publisher = {Springer Nature Singapore},
  address   = {Singapore},
  year      = {2025},
  edition   = {1},
  pages     = {XIX, 484},
  isbn      = {978-981-96-8200-3},
  doi       = {10.1007/978-981-96-8200-3},
  url       = {https://doi.org/10.1007/978-981-96-8200-3}
}

@article{Guastoni2021,
  author = {Guastoni, Luca and Encinar, M. P. and Schlatter, Philipp and Azizpour, Hossein and Vinuesa, Ricardo},
  title = {Convolutional-network Models to Predict Wall-bounded Turbulence from Wall Quantities},
  journal = {Journal of Fluid Mechanics},
  volume = {928},
  pages = {A27},
  year = {2021},
  doi = {10.1017/jfm.2021.812}
}

@article{Lu2021DeepONet,
  author = {Lu, Lu and Jin, Pengzhan and Pang, Guofei and Zhang, Zhongqiang and Karniadakis, George Em},
  title = {Learning Nonlinear Operators via DeepONet Based on the Universal Approximation Theorem of Operators},
  journal = {Nature Machine Intelligence},
  volume = {3},
  pages = {218--229},
  year = {2021},
  doi = {10.1038/s42256-021-00302-5}
}

@article{Raissi2019,
  author = {Raissi, Maziar and Perdikaris, Paris and Karniadakis, George Em},
  title = {Physics-informed Neural Networks: A Deep Learning Framework for Solving Forward and Inverse Problems Involving Nonlinear Partial Differential Equations},
  journal = {Journal of Computational Physics},
  volume = {378},
  pages = {686--707},
  year = {2019},
  doi = {10.1016/j.jcp.2018.10.045}
}

@article{Karniadakis2021,
  author = {Karniadakis, George Em and Kevrekidis, Ioannis G. and Lu, Lu and Perdikaris, Paris and Wang, Sifan and Yang, Liu},
  title = {Physics-informed Machine Learning},
  journal = {Nature Reviews Physics},
  volume = {3},
  pages = {422--440},
  year = {2021},
  doi = {10.1038/s42254-021-00314-5}
}

@article{Kovachki2023,
  author = {Kovachki, Nikola and Li, Zongyi and Liu, Burigede and Azizzadenesheli, Kamyar and Bhattacharya, Kaushik and Stuart, Andrew M. and Anandkumar, Anima},
  title = {Neural Operator: Learning Maps Between Function Spaces with Applications to PDEs},
  journal = {Acta Numerica},
  volume = {32},
  pages = {1--97},
  year = {2023},
  doi = {10.1017/S0962492923000051}
}

@article{Brunton2020,
  author = {Brunton, Steven L. and Noack, Bernd R. and Koumoutsakos, Petros},
  title = {Machine Learning for Fluid Mechanics},
  journal = {Annual Review of Fluid Mechanics},
  volume = {52},
  pages = {477--508},
  year = {2020},
  doi = {10.1146/annurev-fluid-010719-060214}
}

@article{Duraisamy2019,
  author = {Duraisamy, Karthik and Iaccarino, Gianluca and Xiao, Heng},
  title = {Turbulence Modeling in the Age of Data},
  journal = {Annual Review of Fluid Mechanics},
  volume = {51},
  pages = {357--377},
  year = {2019},
  doi = {10.1146/annurev-fluid-010518-040547}
}

@article{Kochkov2021,
  author = {Kochkov, Dmitrii and Smith, Jamie A. and Alieva, Ayya and Wang, Qing and Brenner, Michael P. and Hoyer, Stephan},
  title = {Machine Learning--Accelerated Computational Fluid Dynamics},
  journal = {Proceedings of the National Academy of Sciences},
  volume = {118},
  pages = {e2101784118},
  year = {2021},
  doi = {10.1073/pnas.2101784118}
}

@article{Xiao2023,
  author = {Xiao, Tianbai and Frank, Martin},
  title = {RelaxNet: A Structure-preserving Neural Network to Approximate the Boltzmann Collision Operator},
  journal = {Journal of Computational Physics},
  volume = {490},
  pages = {112317},
  year = {2023},
  doi = {10.1016/j.jcp.2023.112317}
}

@article{Miller2022,
  author = {Miller, Scott T. and Roberts, Nathan V. and Cyr, Eric C.},
  title = {Neural-network Based Collision Operators for the Boltzmann Equation},
  journal = {Journal of Computational Physics},
  volume = {470},
  pages = {111541},
  year = {2022},
  doi = {10.1016/j.jcp.2022.111541}
}

@article{Corbetta2023,
  author = {Corbetta, Alessandro and van den Eijnden, Erik and Toschi, Federico},
  title = {Toward Learning Lattice Boltzmann Collision Operators},
  journal = {The European Physical Journal E},
  volume = {46},
  pages = {20},
  year = {2023},
  doi = {10.1140/epje/s10189-023-00264-6}
}

@article{Ball2025,
  author = {Ball, Nicholas Daultry and MacArt, Jonathan F. and Sirignano, Justin},
  title = {Online Optimisation of Machine Learning Collision Models to Accelerate Direct Molecular Simulation of Rarefied Gas Flows},
  journal = {Journal of Computational Physics},
  volume = {524},
  pages = {113681},
  year = {2025},
  doi = {10.1016/j.jcp.2025.113681}
}

@article{RoohiAST2026,
  author = {Roohi, Ehsan and Shoja-Sani, Ahmad},
  title = {Data-driven Surrogate Modeling of DSMC Solutions Using Deep Neural Networks},
  journal = {Aerospace Science and Technology},
  volume = {168},
  pages = {110785},
  year = {2026},
  doi = {10.1016/j.ast.2025.110785}
}

@article{RoohiPoF2026a,
  author = {Roohi, Ehsan and Shoja-Sani, Ahmad and Ebrahimzadeh Azghadi, Fahimeh},
  title = {Neural Networks for Rarefied Gas Dynamics: Relaxation Problem, Polyatomic Shock Waves, and Hypersonic Cylinder Flow},
  journal = {Physics of Fluids},
  volume = {38},
  pages = {057108},
  year = {2026},
  doi = {10.1063/5.0334590}
}

@article{RoohiPoF2026b,
  author = {Roohi, Ehsan and Shoja-Sani, Ahmad and Stefanov, Stefan},
  title = {Physics Constrained Neural Collision Operators for Hard Sphere Surrogates and Ab Initio Angle Prediction in Direct Simulation Monte Carlo},
  journal = {Physics of Fluids},
  volume = {38},
  pages = {057123},
  year = {2026},
  doi = {10.1063/5.0328463}
}

@article{RoohiMicro2026,
  author = {Roohi, Ehsan and Mahdavi, Amirmehran},
  title = {Analysis of the Rarefied Flow at Micro-step Using a DeepONet Surrogate Model with a Physics-guided Zonal Loss Function},
  journal = {Microfluidics and Nanofluidics},
  volume = {30},
  pages = {44},
  year = {2026},
  doi = {10.1007/s10404-026-02899-8}
}

@article{SharipovStrapasson2012,
  author = {Sharipov, Felix and Strapasson, Jos\'{e} L.},
  title = {Ab Initio Simulation of Transport Phenomena in Rarefied Gases},
  journal = {Physical Review E},
  volume = {86},
  pages = {031130},
  year = {2012},
  doi = {10.1103/PhysRevE.86.031130}
}

@article{SharipovStrapasson2013,
  author = {Sharipov, Felix and Strapasson, Jos\'{e} L.},
  title = {Benchmark Problems for Mixtures of Rarefied Gases. I. Couette Flow},
  journal = {Physics of Fluids},
  volume = {25},
  pages = {027101},
  year = {2013},
  doi = {10.1063/1.4791604}
}

@article{StrapassonSharipov2014Heat,
  author = {Strapasson, Jos\'{e} L. and Sharipov, Felix},
  title = {Ab Initio Simulation of Heat Transfer Through a Mixture of Rarefied Gases},
  journal = {International Journal of Heat and Mass Transfer},
  volume = {71},
  pages = {91--97},
  year = {2014},
  doi = {10.1016/j.ijheatmasstransfer.2013.12.011}
}

@article{SharipovDias2017Shock,
  author = {Sharipov, Felix and Dias, Fernanda C.},
  title = {Ab Initio Simulation of Planar Shock Waves},
  journal = {Computers \& Fluids},
  volume = {150},
  pages = {115--122},
  year = {2017},
  doi = {10.1016/j.compfluid.2017.04.002}
}

@article{Sharipov2022DSMC,
  author = {Sharipov, Felix},
  title = {Direct Simulation Monte Carlo Method Based on Ab Initio Potential: Recovery of Transport Coefficients of Multi-component Mixtures of Noble Gases},
  journal = {Physics of Fluids},
  volume = {34},
  pages = {097114},
  year = {2022},
  doi = {10.1063/5.0106661}
}

\end{document}